\documentclass[a4paper,11pt]{article}
\pdfoutput=1
\usepackage{jheppub} 
\usepackage{graphicx}
\usepackage{amsmath}
\usepackage{amssymb}
\usepackage{bbold}
\usepackage{float}
\usepackage{slashed}
\usepackage{dcolumn}% Align table columns on decimal point
\usepackage{bm}% bold math 
\usepackage{color}
\usepackage{multirow}

\allowdisplaybreaks

\newcommand{\lsim}{{\;\raise0.3ex\hbox{$<$\kern-0.75em\raise-1.1ex\hbox{$\sim$}}\;}}
\newcommand{\gsim}{{\;\raise0.3ex\hbox{$>$\kern-0.75em\raise-1.1ex\hbox{$\sim$}}\;}}
\def\bea{\begin{eqnarray}}
\def\eea{\end{eqnarray}}
\def\bec{\begin{center}}
\def\ec{\end{center}}

\def\beq{\begin{equation}}
\def\eeq{\end{equation}}

\def\bea{\begin{eqnarray}}
\def\eea{\end{eqnarray}}
\def\beq#1\eeq{\begin{align}#1\end{align}}
\def\beqnn#1\eeq{\begin{align*}#1\end{align*}}
\def\ba{\begin{array}}
\def\ea{\end{array}}
\def\bc{\begin{center}}
\def\ec{\end{center}}

\newcommand{\dis}[1]{\begin{equation}\begin{split}#1\end{split}\end{equation}}

\preprint{CTPU-PTC-23-34}

\title{Exploring CP Violation beyond the Standard Model  \\ and
the PQ Quality  with Electric Dipole Moments}
 
 \author{
    Kiwoon Choi\footnote{email: kchoi@ibs.re.kr},
    Sang Hui Im\footnote{email: imsanghui@ibs.re.kr}, 
    Krzysztof Jod\l{}owski\footnote{email: k.jodlowski@ibs.re.kr}
 }
     
\affiliation{
Particle Theory and Cosmology Group, Center for Theoretical Physics of the Universe, \\
 Institute for Basic Science (IBS), Daejeon 34126, Korea \
    }

\abstract{
In some models of physics beyond the Standard Model (SM), one of the leading low energy consequences of the model  appears in the form of  the chromo-electric dipole moments (CEDMs) of the gluons and light quarks. We  examine if  these CEDMs can be distinguished from the QCD $\theta$-term through the experimentally measurable nuclear and atomic electric dipole moments (EDMs)
in both cases with and without the Peccei-Quinn (PQ) mechanism solving the strong CP problem. 
We find that the nucleon EDMs show a distinctive pattern when the EDMs are dominantly induced by the light quark CEDMs without the PQ mechanism. 
In the presence of the PQ mechanism, the QCD $\theta$-parameter corresponds to the vacuum value of the axion field, which might be induced either by CEDMs or  by
UV-originated PQ breaking other than the QCD anomaly, for instance the PQ breaking by quantum gravity effects. We find that in case with the PQ mechanism the nucleon EDMs  have a similar pattern regardless of what is the dominant source of EDMs among the CEDMs and $\theta$-term, unless there is a significant cancellation between the contributions from different sources.
In contrast, some nuclei or atomic EDMs can have characteristic patterns significantly depending on the dominant source  of EDMs, which may allow identifying the dominant source  among the CEDMs and $\theta$-term.
Yet, discriminating the gluon CEDM from the QCD $\theta$-parameter necessitates additional knowledge of low energy parameters induced by the gluon CEDM, which is not available at the moment.
Our results imply that   EDMs can reveal unambiguous sign of CEDMs while identifying the origin of the axion vacuum value, however it requires further knowledge of low energy parameters  induced by the gluon CEDM.
}

\vspace{4cm}

\begin{document} 
\maketitle
\flushbottom
 
\section{Introduction}

Permanent electric dipole moments (EDM) of particles are known to provide a sensitive tool to probe CP violation beyond the Standard Model (SM) of particle physics.  Furthermore, the sensitivity of the experimental search for EDMs is expected to be significantly improved  within the foreseeable future (see e.g. \cite{2203.08103}).  As is well known, CP violation in the SM  (up to dim $=4$ operators) can be described by the two angle parameters,
the Kobayashi-Maskawa phase $\delta_{\rm KM}$ inducing CP violation in the weak interactions and the QCD angle
$\bar\theta$ for CP violation in the strong interactions. These two angle parameters are determined by the SM parameters as \cite{Jarlskog:1985cw, Jarlskog:1985ht}
\bea
\delta_{\rm KM}={\rm arg}\cdot{\rm det}([Y_u Y_u^\dagger, Y_d Y_d^\dagger]), \quad \bar\theta=\theta_0+{\rm arg}\cdot{\rm det}(Y_u Y_d), \label{cpv_sm}\eea
where $Y_u$ and $Y_d$ are the complex  Yukawa couplings of the 3 generations of the up-type and down-type quarks,
and $\theta_0$ is the bare QCD angle.   CP violating phenomena associated with $\delta_{\rm KM}$ have been 
experimentally well tested,  implying $\delta_{\rm KM}={\cal O}(1)$ \cite{ParticleDataGroup:2022pth}. On the other hand, CP violation by $\bar\theta$ in the strong interactions is not observed yet, which results in the stringent upper bound \cite{Abel:2020pzs, Baluni:1978rf, Crewther:1979pi, Pospelov:1999ha, Chan:1997fw} 
\bea
|\bar\theta|\,\lesssim\, 10^{-10}.\label{strong_cp}\eea

Although $\delta_{\rm KM}$ is of order unity, EDMs induced by  $\delta_{\rm KM}$ are highly suppressed by the involved quark masses and mixing parameters \cite{Pospelov:2013sca}.
As a result they all have a value well below the current experimental bounds. On the other hand,  $\bar\theta$ can generate hadronic EDMs near the current bound, if $\bar\theta$ has a value near $10^{-10}$.
Generically there can also be CP-violating interactions beyond the SM (BSM), which may result in EDMs again near the current experimental bounds. Therefore, once a nonzero hadronic EDM were detected experimentally, one of the key questions is whether it originates from $\bar\theta$ or  from BSM CP violation. To answer this question, one needs to measure multiple EDMs
in the experimental side, and examine in the theory side if the observed pattern of EDMs can be explained by $\bar\theta$ or requires an alternative source of CP violation. Previous studies along this line include \cite{Lebedev:2004va, Dekens:2014jka, deVries:2018mgf, deVries:2021sxz}. Generically one may consider an effective theory defined at a scale around the QCD scale,  involving the flavor-conserving CP-odd effective interactions
\bea
\Delta {\cal L} = \frac{g_s^2}{32\pi^2}\bar\theta G^{a\mu\nu}\tilde G^a_{\mu\nu}+\sum_i \lambda_i {\cal O}_i, \label{effective_cp_odd}
\eea
and examine which region of the parameter space of $\{\bar\theta, \lambda_i\}$ can explain the observed pattern of EDMs,
where ${\cal O}_i$ denote the non-renormalizable ($\mbox{dim} > 4$)  flavor-conserving CP-odd local operators of light fields,
which would describe the low energy consequence of generic BSM physics existing at higher energy scales,
and $\lambda_i$ are their Wilson coefficients.  
 Such  effective interactions then include 
the chromo-electric dipole moments (CEDMs) of the gluons and light quarks, EDMs of the light quarks and leptons, and various forms of four-fermion operators (see e.g. \cite{Dekens:2014jka, deVries:2021sxz}).

In view of the expression  (\ref{cpv_sm}) for  $\delta_{\rm KM}$ and $\bar\theta$,
the smallness of $\bar\theta$ requires a severe fine-tuning. An appealing solution to this problem is to introduce
a global $U(1)$ Peccei-Quinn (PQ) symmetry \cite{Peccei:1977hh, Weinberg:1977ma, Wilczek:1977pj} (see e.g. \cite{0807.3125, 2003.01100, Choi:2020rgn}  for reviews)
which is non-linearly realized at least in  low energy limits, under which the associated Nambu-Goldstone boson, the axion $a(x)$, transforms as
\bea
U(1)_{\rm PQ}: \,\,\, a(x)\rightarrow a(x)+{\rm constant}. \label{u1_pq}\eea
A key assumption involved in this solution is that $U(1)_{\rm PQ}$ is broken {\it dominantly} by the QCD anomaly,  i.e. by the axion coupling to the gluons of the form
\bea
\frac{g_s^2}{32\pi^2}\frac{a(x)}{f_a} G^{a\mu\nu}\tilde G^a_{\mu\nu}\label{axion_gluon} \label{axion_gluon},
\eea
to the extent that the resulting axion vacuum value is small enough to satisfy
\bea
\bar\theta = \frac{\langle a(x)\rangle}{f_a} \lesssim 10^{-10}.
\eea

Yet, the PQ mechanism does not predict the value of $\bar\theta=\langle a\rangle/f_a$. Generically there can be a variety of model-dependent physics generating
nonzero axion vacuum value, which may give any value of $\bar\theta$ below $10^{-10}$.  It includes for instance (i) BSM physics generating the CP-odd effective interactions
$\sum_i\lambda_i{\cal O}_i$ in (\ref{effective_cp_odd}), which would shift the axion vacuum value when it is combined with
the  $U(1)_{\rm PQ}$-breaking by the QCD anomaly, as well as (ii) additional, typically UV-originated, $U(1)_{\rm PQ}$-breaking {\it other than} the QCD anomaly, e.g. quantum gravity effects,  which would by itself generate an axion potential at the corresponding UV scale.   EDMs then may provide a way to discriminate these two potentially dominant origins of the axion vacuum value from each other, since the origin (i) affects EDMs both directly and through the induced axion vacuum value, while the origin (ii) affects EDMs mostly through the induced axion vacuum value.
This suggests that EDMs can provide information not only on BSM CP violation, but also 
{on the origin of the axion vacuum value, therefore  on the quality of the PQ symmetry characterized by the strength of UV-originated $U(1)_{\rm PQ}$-breaking other than the 
QCD anomaly.}

In this paper, we examine if certain class of BSM CP violations can give rise to a distinguishable pattern of nucleon and atomic EDMs from the pattern due to $\bar\theta$, in both cases with and without the PQ mechanism. 
We also examine if these EDMs can identify the origin of the axion vacuum value, specifically if they can discriminate the axion vacuum value $\bar\theta_{\rm UV}=\langle a\rangle_{\rm UV}/f_a$ 
induced by the origin (ii) from $\bar\theta_{\rm BSM}=\langle a\rangle_{\rm BSM}/f_a$ induced by the origin (i).
For simplicity, we focus on BSM CP violation  whose low energy consequence appears mainly  in the form of the gluon and quark CEDMs around the weak scale.

We first find that the nucleon EDMs  show a distinctive pattern when the EDMs are dominantly induced by the light quark CEDMs without the PQ mechanism. 
In the presence of the PQ mechanism, CEDMs need to be compared with
 $\bar\theta_{\rm UV}$.
 We  then find that 
the nucleon EDMs due to the gluon or light quark CEDMs
in the presence of the PQ mechanism
  have a similar pattern as those due to $\bar\theta_{\rm UV}$,
  unless there is a significant cancellation between the contributions from different sources.
   Note that in this case the EDMs due to CEDMs include the contributions from $\bar\theta_{\rm BSM}$ induced by CEDMs. 
In contrast,  {some} light nuclei and atomic EDMs due to the light quark CEDMs have characteristic patterns  distinguishable from the pattern due to $\bar\theta_{\rm UV}$.  
Yet those EDMs can not unambiguously distinguish the gluon CEDM from $\bar\theta_{\rm UV}$, mainly due to the lack of 
knowledge about low energy parameters induced by the gluon CEDM.  
Our results imply that   EDMs can provide an unambiguous sign of BSM CP violation while identifying the origin of the axion vacuum value, however it requires further knowledge on low energy parameters  associated with BSM CP violation.

The organization of this paper is as follows. In the next section, we briefly discuss the quality of the PQ symmetry which is  about the axion vacuum value  in the presence of both
BSM CP violation and  UV-originated $U(1)_{\rm PQ}$ breaking other than the QCD anomaly.  In section \ref{sec:GH},  we discuss BSM CP violation mediated mainly by the SM gauge bosons and/or the Higgs boson, as well as the resulting CEDMs 
of the gluons and light quarks at low energy scales. Section \ref{sec:EDM} is devoted to the analysis of nuclear and atomic EDMs induced by $\bar\theta$ and the gluon and quark CEDMs in both cases with and without the PQ mechanism. In section \ref{sec:examples}, we provide some examples of BSM models yielding
low energy CP violations dominated by the gluon and light quark CEDMs. Section \ref{sec:conc} is the conclusion.

\section{PQ quality with BSM CP violation} \label{sec:PQ}

In models with QCD axion, the axion potential is generically given by
\bea
V(a) = V_{\rm QCD}(a) +\delta V(a)
\eea
where \bea
V_{\rm QCD}(a)\simeq -\frac{f_\pi^2 m_\pi^2}{(m_u+m_d)}\sqrt{m_u^2+m_d^2+2m_um_d \cos(a/f_a)}
\eea
is the axion potential induced by the $U(1)_{\rm PQ}$-breaking by the QCD anomaly \cite{Choi:2020rgn},  i.e.  the axion coupling  (\ref{axion_gluon}), which has the global minimum\footnote{Here the axion field is defined in such a way that $\langle a\rangle/f_a$ is identified as the QCD angle $\bar\theta$ violating CP in the strong interactions, which can be always done by an appropriate constant shift of the axion field.} at $\langle a\rangle =0$, and $\delta V$ denotes the model-dependent additional axion potential
which has a minimum  at $\langle a\rangle \neq0$.
Here $m_{u,d}$ are the light quark masses.

Generically there can be two different sources of  $\delta V$.  The first is the {\it combined} effect of the PQ-breaking by the QCD anomaly and a CP violating effective interaction of gluons and/or light quarks  around the QCD scale where the QCD anomaly becomes important. 
This includes, first of all, the SM contribution \cite{Georgi:1986kr} \bea\delta V_{\rm SM} \sim 10^{-19} f_\pi^2 m_\pi^2 \sin\delta_{\rm KM}\sin(a/f_a),\eea which results in 
\bea
\bar\theta_{\rm SM} =\frac{\langle a\rangle_{\rm SM}}{f_a} \sim 10^{-19}\sin\delta_{\rm KM},\eea
which is too small to be phenomenologically interesting in the near future.
On the other hand, in the presence of BSM physics generating CP-odd effective interactions around the QCD scale, the resulting $\bar\theta$ might be as large as $10^{-10}$. For instance, for the effective interactions given by
\bea
 {\cal L}_{\rm eff}=\sum_i \lambda_i {\cal O}_i,\label{effective_op}
\eea
where ${\cal O}_i$ are non-renormalizable flavor-conserving CP-odd effective interactions of  the gluons and/or light quarks and $\lambda_i$ are the associated  Wilson coefficients, one finds
\bea
\left. f_a  \frac{\partial \delta V_{\rm BSM}}{\partial a}\right|_{a=0} \sim\,\sum_i \lambda_i  \int d^4 x \,\Big\langle\frac{g_s^2}{32\pi^2}G\tilde G(x) {\cal O}_i(0)\Big\rangle_{a=0}.
\eea
The resulting shift of the axion vacuum value is given by
\bea
 \bar\theta_{\rm BSM}=\frac{\langle a\rangle_{\rm BSM}}{f_a} \sim {\frac{\sum_i \lambda_i  \int d^4 x \,\Big\langle\frac{g_s^2}{32\pi^2}G\tilde G(x) {\cal O}_i(0)\Big\rangle_{a=0}}{f_\pi^2 m_\pi^2}} \label{theta_bsm}
\eea
which can have any value  below $10^{-10}$. 
 
The second potentially dominant source of $\delta V$ is additional, typically UV-originated, PQ breaking other than the QCD anomaly.  For instance, it has been argued that generically quantum gravity does not respect global symmetries, so can generate a $U(1)_{\rm PQ}$-breaking axion potential  around the scale of quantum gravity \cite{Barr:1992qq,  Kamionkowski:1992mf, Holman:1992us, Ghigna:1992iv}.  Study of axions in string theory and also of axionic Euclidean wormholes imply that string/brane instantons or gravitational wormholes generate (For reviews, see for instance \cite{0902.3251, 1807.00824}.) 
\bea
\delta V_{\rm UV} = \Lambda_{\rm UV}^4 e^{-S_{\rm ins}}\cos(a/f_a +\delta_{\rm UV}) \label{instanton_po},
\eea
where $\Lambda_{\rm UV}$ is a model-dependent UV scale\footnote{Often it is given by $\Lambda_{\rm UV}^4\sim m_{3/2}M_{\rm Pl}^3$ or $m_{3/2}^2M_{\rm Pl}^2$ \cite{Dine:1986zy, hep-ph/9902292, 0902.3251}
for axions in string theory, where $M_{\rm Pl}\simeq 2\times 10^{18}$ GeV is the reduced Planck scale and $m_{\rm 3/2}$ is the gravitino mass.}, $S_{\rm ins}$ is the Euclidean action of the associated string/brane instanton or of the Euclidean wormhole, and $\delta_{\rm UV}$ is a phase angle which is generically of order unity.
This shifts the axion vacuum value as 
\bea
 \bar\theta_{\rm UV}=\frac{\langle a\rangle_{\rm UV}}{f_a} \sim {e^{-S_{\rm ins}} \Lambda_{\rm UV}^4\sin\delta_{\rm UV} }/{f_\pi^2 m_\pi^2} \label{thetaUV}
\eea
which again can have any value  below $10^{-10}$.

As noted in the previous section, the above two origins of nonzero axion vacuum value may give distinguishable patterns of EDMs because  the effective interaction (\ref{effective_op}) affects
EDMs both directly and through the induced axion vacuum value, while the additional $U(1)_{\rm PQ}$ breaking generating the axion potential (\ref{instanton_po}) affects EDMs mostly through the induced axion vacuum value.  
As we will see, for the case that BSM CP violation around the QCD scale is dominated by the gluon and light quark CEDMs, the two origins can give distinguishable patterns of EDMs.

\section{BSM CP violation mediated by the SM gauge and Higgs bosons} \label{sec:GH}

As for BSM CP violation, for simplicity, our analysis focuses on a class of BSM scenarios in which the new physics sector communicates with the SM sector dominantly through the SM gauge interactions and/or the couplings to the SM Higgs boson,  which has been dubbed ``universal" theories \cite{hep-ph/0405040}. The new physics sector can generally involve CP-violating (CPV) interactions.
In such cases integrating out heavy fields of the new physics sector would give rise to CP-odd dimension-six operators composed of the SM gauge fields and the Higgs field as follows,
\dis{
{\cal L}_{\rm CPV} (\mu = \Lambda)=&c_{\widetilde G} f^{abc} G_{\alpha}^{a \mu}  G_\mu^{b\delta} \widetilde{G}_\delta^{c\alpha}+ c_{\widetilde W}  \epsilon^{abc} W_{\alpha}^{a \mu}  W_\mu^{b\delta} \widetilde{W}_\delta^{c\alpha} \\
&+ |H|^2 \left( c_{H\widetilde{G}} G^a_{\mu \nu} \widetilde{G}^{a \mu \nu} + c_{H\widetilde{W}}   W^a_{\mu \nu} \widetilde{W}^{a \mu \nu} +c_{H\widetilde{B}} B_{\mu \nu} \widetilde{B}^{\mu \nu}\right) \\
&+ c_{H\widetilde{W} B} H^\dagger \tau^a H \widetilde{W}^a_{\mu \nu} B^{\mu \nu} \label{eft1}
}
with the Wilson coefficients $c_i$ defined at a certain scale $\mu = \Lambda$ which is around the mass scale of the heavy fields in the new physics sector. 
Here $f^{abc}$ is the structure constant for the color gauge group $SU(3)_c$, $\epsilon^{abc}$ is the structure constant for the weak gauge group $SU(2)_W$, and $\tau^a$ is the Pauli matrix for $SU(2)_W$.
At one-loop level the following operators can also be induced either directly or by the renormalization group evolution (RGE) of the operators in Eq. (\ref{eft1}):
\dis{
\sum_{q=u, d}  \sum_{X=G, W, B} i(c_{q X})_{ij} \bar{Q}_{Li} \sigma^{\mu \nu} X_{\mu \nu} q_{Rj} H^{(*)} + \sum_{X=W, B} i(c_{e X})_{ij} \bar{L}_{i} \sigma^{\mu \nu} X_{\mu \nu} e_{Rj} H^{(*)} + \textrm{h.c.}   \label{eft1-1},
} 
where $i, j$ are flavor indices, and $H^{(*)}\equiv H $ or $H^*$ in order to make the operators invariant under the SM gauge group. {The CP violating phenomenology of the lagrangian (\ref{eft1}) and (\ref{eft1-1}) was first analyzed in \cite{1903.03625}}. The full one-loop RG equations of the involved operators over the scales between the BSM scale and the electroweak scale are given in appendix \ref{app:RGE} using the results of \cite{1303.3156, Jenkins:2013zja, Jenkins:2013wua, Alonso:2013hga}. Here we only show the dominant RGE effect involving the QCD coupling and the flavor-diagonal part:
\dis{
16\pi^2\frac{d c_{\widetilde G}}{d \ln \mu} &=  (N_c + 2n_F) g_s^2 c_{\widetilde G}\,,\\
16\pi^2  \frac{d (c_{qG})_{ii}}{d \ln \mu} 
&=- \left(\frac83 N_c +\frac{5}{N_c} -\frac23 n_F \right) g_s^2 (c_{qG})_{ii} + (Y_q)_{ii} \left(-4g_s  c_{H \widetilde G} +3 N_c g_s^2  c_{\widetilde G} \right), \\
16\pi^2 \frac{d c_{H\widetilde  G}}{d\ln\mu} &=-\frac{2}{3}(11N_c -2n_F)g_s^2  c_{H\widetilde G}+\left( 2 i g_s  \textrm{Tr}[Y_u c_{uG} + Y_d c_{dG}]  + \textrm{h.c.} \right), \label{RGuv}  
}
where $N_c=3$ is the number of the QCD color, $n_F=6$ is the number of the Dirac quarks, and $(Y_q)_{ii}$ is the flavor-diagonal quark Yukawa coupling. 

Below the electroweak scale the Higgs field and $W/Z$-field are integrated out. Consequently, the leading effective CPV interactions from the operators in Eq. (\ref{eft1}) and Eq. (\ref{eft1-1}) are given by
\dis{
{\cal L}_{\rm CPV} (\mu = m_W) &=\frac{1}{3} w f^{abc} G_{\alpha}^{a \mu}  G_\mu^{b\delta} \widetilde{G}_\delta^{c\alpha}
-\frac{i}{2} \sum_q \tilde{d}_q g_s \bar{q} \sigma^{\mu \nu} G_{\mu \nu} \gamma_5 q
-\frac{i}{2} \sum_{f=q, \ell} d_f e \bar{f} \sigma^{\mu \nu} F_{\mu \nu} \gamma_5 f  \\
&+\frac{g_s^2}{32\pi^2}  \theta(m_W) G^a_{\mu \nu} \widetilde{G}^{a \mu \nu}, \label{eft2}
}
where  $\theta(m_W)$ includes the threshold correction from the $c_{H \widetilde{G}}$-term in Eq. (\ref{eft1}), and the Wilson coefficients are determined by the following matching conditions at $\mu = m_W$:
\dis{
\frac13 w&=  c_{\tilde{G}}\,, \\
g_s \tilde{d}_{q_i} &=  \sqrt{2} v (c_{qG})_{ii}\,, \\
e\, d_{f_i} &= \sqrt{2} v (s_w c_{f W} + c_w c_{f B})_{ii}\,.
}
Here $v=246$ GeV, $s_w = \sin \theta_w$, $c_w = \cos \theta_w$ with the weak mixing angle $\theta_w$, and $q$ and $\ell$ stand for active light Dirac quarks and leptons, respectively. 
Thus, the low energy CPV effect mediated by gauge and Higgs interactions is characterized mainly by the gluon chromo-electric dipole moment (CEDM) ($=$ the Weinberg three-gluon operator), quark CEDMs, and quark and lepton electric dipole moments (EDMs). 
On the other hand, the new physics contributions to the QCD $\theta$-parameter would be indistinguishable from the SM bare value. 

Since the lepton EDMs from the SM are predicted to be far below the current experimental bounds 
\cite{Choi:1990cn, Ghosh:2017uqq, Pospelov:2013sca}, we may be able to distinguish BSM CPV from the SM one by the lepton EDMs if the lepton EDMs from new physics are sizable. 
On the other hand, in this work, we will examine whether one can discriminate BSM CPV by means of the hadronic EDMs. 
Furthermore, we will focus on the case that the low energy CPV is characterized mainly by the gluon and quark CEDMs, while the quark EDMs are subdominant. We will discuss in section \ref{sec:examples} that it is typically the case if the lightest new physics sector communicating with the SM through gauge and Higgs interactions carries the QCD color. More general studies including the case that the quark EDMs are potentially dominant source of CPV are subject to future work \cite{preparation}.

The CP violation through the gluon and quark CEDMs will give rise to electric dipole moments of nucleons and atoms below the QCD scale. 
In order to estimate the nucleon and atomic EDMs, we need to bring the Wilson coefficients down to the QCD scale ($\sim$ 1 GeV) through the RGE. 
This running effect is important because the QCD gauge coupling becomes large $(g_s^2/4\pi^2 \sim 1)$ near the QCD scale. 
The relevant RGE equations at leading order are given by \cite{Morozov:1984goy, Braaten:1990gq, Chang:1991hz, Degrassi:2005zd, Hisano:2012cc}
\bea
\frac{d  {\bf K}}{d \ln \mu} &=& \frac{g_s^2}{16\pi^2} \gamma \, {\bf K}, \label{RGECs} 
\eea
where  ${\bf K} \equiv (K_1~ K_2~ K_3)^T$ are defined as 
\bea
K_1(\mu) = \frac{d_q(\mu)}{m_q Q_q}, 
\quad 
K_2(\mu) = \frac{\tilde{d}_q(\mu)}{m_q}, 
\quad 
K_3(\mu) = \frac{w(\mu)}{g_s}, \label{C123}
\eea
and the anomalous dimension matrix $\gamma$ is
\bea
\gamma \equiv
\left(
\begin{array}{ccc}
\gamma_e &\gamma_{eq}  & 0
\\
0 & \gamma_q & \gamma_{Gq}
\\
0 &0 & \gamma_G
\end{array}
\right)
= 
\left( 
\begin{array}{ccc}
8 C_F & 8 C_F & 0 \\
0 & 16C_F -4 N_c &  -2 N_c \\
0 & 0 & N_c + 2 n_f +\beta_0
\end{array}
\right).
\eea
Here $C_F = (N_c^2-1)/2N_c= 4/3$ is the quadratic Casimir, $N_c = 3$ is the number of color, $n_f$ is the number of active light Dirac quarks, and $\beta_0 \equiv (33-2n_f)/3$.
The color fine structure constant $\alpha_s = g_s^2/4\pi$ and the quark mass run according to
\dis{
\frac{d \alpha_s}{d \ln \mu} = - 2\beta_0 \frac{\alpha_s^2}{4\pi}, \quad \frac{d m_q}{d \ln \mu} = -8 \frac{\alpha_s}{4\pi} m_q\,. \label{alphasmq}
}
Using Eq. (\ref{alphasmq}), the analytic solution to the RGE equations is obtained as \cite{Degrassi:2005zd}
\bea
K_1 (\mu) &=&
\eta^{\kappa_e} K_1(\Lambda) + \frac{\gamma_{qe}}{\gamma_e - \gamma_q} (\eta^{\kappa_e} - \eta^{\kappa_q} )K_2(\Lambda)  \nonumber \\
&+& \left[ \frac{\gamma_{Gq}\gamma_{qe} \eta^{\kappa_{e}} }{(\gamma_q - \gamma_e)(\gamma_G -\gamma_e)} 
+ \frac{\gamma_{Gq}\gamma_{qe} \eta^{\kappa_{q}} }{(\gamma_e - \gamma_q)(\gamma_G -\gamma_q)}  
+ \frac{\gamma_{Gq}\gamma_{qe} \eta^{\kappa_{G}} }{(\gamma_e - \gamma_G)(\gamma_q -\gamma_G)} 
\right] K_3(\Lambda),  \nonumber 
\\
K_2(\mu) &=& 
\eta^{\kappa_q}K_2(\Lambda) 
+ \frac{\gamma_{Gq}}{\gamma_q - \gamma_G} 
\left[
\eta^{\kappa_q} - \eta^{\kappa_G} \right] K_3(\Lambda), \nonumber
\\
K_3(\mu) &=& \eta^{\kappa_G} K_3(\Lambda), \label{analytic}
\eea
where $\eta = \alpha_s(\Lambda) / \alpha_s (\mu)$ and $\kappa_x = \gamma_x / (2\beta_0)$. 

For the renormalization scale $\mu < m_c$ and the BSM scale $\Lambda \geq 1$ TeV, we derive the following analytic relations from Eq. (\ref{RGuv}) and Eq. (\ref{analytic}):
\dis{\hskip -2cm
w(\mu) = {\left( \frac{g_s(m_c)}{g_s(\mu)}\right)} \left( \frac{g_s(m_b)}{g_s(m_c)}\right)^{\frac{33}{25}} \left( \frac{g_s(m_t)}{g_s(m_b)} \right)^{\frac{39}{23}}
 \left( \frac{g_s(\Lambda)}{g_s(m_t)}\right)^{\frac{15}{7}} w(\Lambda), \label{wsol}
 }
 \dis{
\Delta \tilde{d}_q(\mu) &= 
 \frac{m_q(\mu)}{g_s(\Lambda)} \Bigg[
  \frac{9}{13} \left\{\left( \frac{g_s(m_c)}{g_s(\mu)}\right)^{\frac{28}{27}}  - \left( \frac{g_s(m_c)}{g_s(\mu)}\right)^{2}  \right\} \left( \frac{g_s(m_b)}{g_s(m_c)}\right)^{\frac{58}{25}} \left( \frac{g_s(m_t)}{g_s(m_b)} \right)^{\frac{62}{23}}
 \left( \frac{g_s(\Lambda)}{g_s(m_t)}\right)^{\frac{22}{7}}   \\
 &+ \frac{3}{5}\left( \frac{g_s(m_c)}{g_s(\mu)}\right)^{\frac{28}{27}}  \left\{\left( \frac{g_s(m_b)}{g_s(m_c)}\right)^{\frac{28}{25}}  - \left( \frac{g_s(m_b)}{g_s(m_c)}\right)^{\frac{58}{25}}  \right\}  \left( \frac{g_s(m_t)}{g_s(m_b)} \right)^{\frac{62}{23}}  \left( \frac{g_s(\Lambda)}{g_s(m_t)}\right)^{\frac{22}{7}}   \\
 &+ \frac{9}{17}\left( \frac{g_s(m_c)}{g_s(\mu)}\right)^{\frac{28}{27}} \left( \frac{g_s(m_b)}{g_s(m_c)} \right)^{\frac{28}{25}}
   \left\{\left( \frac{g_s(m_t)}{g_s(m_b)}\right)^{\frac{28}{23}}  - \left( \frac{g_s(m_t)}{g_s(m_b)}\right)^{\frac{62}{23}}  \right\}  \left( \frac{g_s(\Lambda)}{g_s(m_t)}\right)^{\frac{22}{7}}   \\
 & +   \frac{9}{19} \left( \frac{g_s(m_c)}{g_s(\mu)}\right)^{\frac{28}{27}} \left( \frac{g_s(m_b)}{g_s(m_c)}\right)^{\frac{28}{25}} \left( \frac{g_s(m_t)}{g_s(m_b)} \right)^{\frac{28}{23}}
 \left\{\left( \frac{g_s(\Lambda)}{g_s(m_t)}\right)^{\frac{4}{3}}  - \left( \frac{g_s(\Lambda)}{g_s(m_t)}\right)^{\frac{22}{7}}  \right\}\Bigg] w(\Lambda),  \label{dqtsol}
 }
where $\Delta \tilde{d}_q(\mu)$ is the RG-induced contribution to the quark CEDM from the gluon CEDM. 
Numerically the above equations give
\dis{\hskip -2cm
w(1 \textrm{ GeV}) \simeq 0.33  \left( \frac{g_s(\Lambda)}{g_s(1 \textrm{ TeV})}\right)^{\frac{15}{7}} w(\Lambda), \label{wrun}
}
\dis{
\frac{\Delta \tilde{d}_q}{m_q}(1 \textrm{ GeV}) &\simeq \left[0.19 \left(\frac{g_s(\Lambda)}{g_s(1 \textrm{ TeV})}\right)^{\frac{1}{3}}-0.06 \left(\frac{g_s(\Lambda)}{g_s(1 \textrm{ TeV})}\right)^{\frac{15}{7}} \right] w(\Lambda). \label{dqtrun}
}
For instance, for the BSM scale $\Lambda = 1$ TeV or 10 TeV, we obtain the following numerical relations which will be useful later
\dis{
\frac{\Delta \tilde{d}_q}{m_q} (1 \textrm{ GeV}) \simeq  
\begin{cases} 0.41\, w(1 \textrm{ GeV}) & \Lambda= 1 \textrm{ TeV}, \\
0.53\,  w(1 \textrm{ GeV}) & \Lambda= 10 \textrm{ TeV}.
\end{cases} \label{rgeff}
} 

The Wilson coefficients around the QCD scale obtained from the above procedure can be matched to hadronic CPV observables such as nucleon EDMs and CP-odd pion-nucleon interactions by a variety of methods. In the next section, we will discuss the resultant nuclear and atomic EDMs from the gluon and quark CEDMs.

\section{Nuclear and Atomic EDMs} \label{sec:EDM}

We have discussed that the BSM CP violation mediated by gauge and Higgs interactions can appear as the gluon and quark CEDMs below the weak scale.
The QCD $\bar{\theta}$-parameter may be another dominant source of hadronic CP violation. 
In this section we estimate the nuclear and atomic EDMs from those CPV operators by matching conditions around the QCD scale $\sim$ 1 GeV in both cases with and without the PQ mechanism.
We will then examine whether the resultant EDM profiles can tell us the origins of CP violation and the quality of the PQ symmetry.

\subsection{Nucleon EDMs}

The nucleon EDMs from the QCD $\bar{\theta}$-parameter and quark (C)EDMs were computed in \cite{Pospelov:2000bw, Hisano:2012sc, Hisano:2015rna} with QCD sum rules. In this approach the nucleon EDMs are associated with the QCD $\bar{\theta}$-parameter and quark (C)EDMs at the renormalization scale $\mu = 1$ GeV as
\dis{
d_N(\bar{\theta}, \tilde{d}_q, d_q) = -c_0\frac{m_N^3 \langle \bar{q} q \rangle}{\lambda_N^2} \Theta_N (\bar{\theta}, \tilde{d}_q, d_q) ,\quad (N = p, n) \label{dN_CEDM}
}
where $c_0 = 1.8 \times 10^{-2}$, $m_N$ is the nucleon mass, $\langle \bar{q} q \rangle = -(0.225 \,\textrm{GeV})^3$ is the quark condensate, $\lambda_N=-0.0436(131) \, \textrm{GeV}^3$ is the coupling between the physical nucleon state and the corresponding interpolating field in the QCD sum rules approach, and
\dis{
\Theta_p(\bar{\theta}, \tilde{d}_q, d_q) = \,& \chi m_*  \left[ (4 e_u - e_d) \left(  \bar{\theta}-\frac{m_0^2}{2} \frac{\tilde{d}_s}{m_s} \right) +\frac{m_0^2}{2}  (\tilde{d}_u - \tilde{d}_d) \left(\frac{4e_u}{m_d} +\frac{e_d}{m_u} \right) \right]   \\
&+\frac{1}{8}(2\kappa+\xi)(4e_u \tilde{d}_u -e_d \tilde{d}_d) + (4d_u-d_d),   \\
\Theta_n(\bar{\theta}, \tilde{d}_q, d_q) =\,& \chi m_*  \left[ (4 e_d - e_u) \left(  \bar{\theta}-\frac{m_0^2}{2} \frac{\tilde{d}_s}{m_s} \right) +\frac{m_0^2}{2}  (\tilde{d}_d - \tilde{d}_u) \left(\frac{4e_d}{m_u} +\frac{e_u}{m_d} \right) \right]  \\
&+\frac{1}{8}(2\kappa+\xi)(4e_d \tilde{d}_d -e_u \tilde{d}_u) + (4d_d-d_u). \label{thepn}
}
Here $m_* \equiv (\sum_{q=u,d,s} m_q^{-1})^{-1}\simeq m_um_d/(m_u+m_d)$, and $e_q$ denotes the electromagnetic (EM) charge of the quark $q$. 
We have also the various susceptibilities of quark condensates defined as \cite{Pospelov:2000bw}:
\dis{
 \langle \bar{q} \sigma_{\mu \nu} q \rangle = e_q \chi F_{\mu \nu} \langle \bar{q} q \rangle,\quad  g_s\langle \bar{q} G_{\mu \nu} q \rangle = e_q \kappa F_{\mu \nu} \langle \bar{q} q \rangle, \\
  g_s \langle \bar{q} G^{\mu \nu} \sigma_{\mu \nu} q \rangle = m_0^2 \langle \bar{q} q \rangle, \quad 2g_s\langle \bar{q} \gamma_5 \tilde{G}_{\mu \nu} q \rangle = i e_q \xi F_{\mu \nu} \langle \bar{q} q \rangle,
}
whose values are given as 
\dis{
&\chi=-5.7(6) \, \textrm{GeV}^{-2}, \,\,\, m_0^2=0.8(1) \, \textrm{GeV}^2, \\
 &\kappa=-0.34(10), \,\,\,  \xi=-0.74(20).
}

On the other hand, the gluon CEDM (Weinberg operator) contribution to the nucleon EDMs was first evaluated in  \cite{Weinberg:1989dx}  by Naive Dimensional Analysis (NDA) as $d_N(w) \approx {\cal O}(e f_\pi w)$ for $w$ defined at the matching scale $\mu_* \simeq 225$ MeV which has been chosen by the condition $\alpha_s(\mu_*)/4\pi \simeq 1/6$ for which the one loop QCD beta function is comparable to the two loop QCD beta function. Later QCD sum rules were used in \cite{Demir:2002gg, Haisch:2019bml}  to compute the one-particle reducible contribution which is obtained by the chiral rotation of the CP-odd nucleon mass. This contribution is proportional to the nucleon \textit{anomalous} magnetic moment $\mu^\mathrm{an}_N$ as\footnote{As it arises from chiral rotation, this one-particle reducible contribution is induced only by the chirality violating anomalous part of the nucleon magnetic moment. We thank Nadoka Yamanaka for useful correspondence on this point.} 
\bea
d_N(w) = -\mu^\mathrm{an}_N\, \frac{3 g_s m_0^2}{32\pi^2}\, w  \ln \frac{M^2}{\mu_{\rm IR}^2}\quad (N = p, n), \label{dN_w}
\eea
where $g_s$ and $w$ are defined at
 $\mu = 1$ GeV, $M$ is the Borel mass in the sum-rule calculation, and $\mu_{\rm IR}$ is the IR cut-off. 
The theoretical uncertainty of the above result
 is about 50\% which originates mostly from $M/\mu_{\rm IR}$ which is taken to be $\sqrt{2} \leq M/\mu_{\rm IR} \leq 2\sqrt{2}$ in \cite{Haisch:2019bml}. Inserting the experimentally measured nucleon anomalous magnetic moments  \cite{ParticleDataGroup:2022pth}
\dis{
  \mu^{\mathrm{an}}_p = (2.79-1)\frac{e}{2m_p} =  1.79 \frac{e}{2m_p}, \,\,\,  \mu^{\mathrm{an}}_n = -1.91 \frac{e}{2m_p}, 
}
which correspond to the respective anomalous magnetic moments with the ``bare'' magnetic moments subtracted, one finds the above one-particle reducible contribution Eq. (\ref{dN_w}) is about factor two  smaller than the NDA estimation.
Recently the one-particle irreducible contribution was also calculated in \cite{Yamanaka:2020kjo} by the non-relativistic quark model, which is about 5 times smaller than the one-particle reducible contribution with the opposite sign. Including both contributions \cite{Yamanaka:2020kjo}, the sum rule calculation leads to
\dis{
d_p(w) =  -18\, w\, e\,\textrm{MeV},   \quad
d_n(w) =  20\, w\,e \,\textrm{MeV}  \label{NEDMw} 
}
with about 60\%  uncertainty, where $w$ is defined at  $\mu = 1$ GeV.

Using the central values for the involved parameters, we obtain from Eq. (\ref{dN_CEDM}) and Eq. (\ref{NEDMw})
\dis{
d_p(\bar{\theta}, \tilde{d}_q, d_q,  w) = \,& -0.46 \times 10^{-16} \bar{\theta} \,e\, \textrm{cm} 
 +  e \left( -0.17  \tilde{d}_u +0.12  \tilde{d}_d + 0.0098 \tilde{d}_s \right) \\
 &+  0.36  d_u -0.09  d_d  -{18} w\, e\,\textrm{MeV},   \\
d_n(\bar{\theta}, \tilde{d}_q, d_q, w) = \,& 0.31 \times 10^{-16} \bar{\theta} \,e\, \textrm{cm}+  e \left( -0.13  \tilde{d}_u +0.16  \tilde{d}_d - 0.0066 \tilde{d}_s \right) \\
&  -0.09 d_u +0.36  d_d+{20} w\, e\,\textrm{MeV}. \label{NEDM} 
}

If the strong CP problem is resolved by the PQ mechanism,
$\bar{\theta}$ is no longer a constant parameter, but the vacuum expectation value (VEV) of the QCD axion, which  is not independent of
hadronic CPV operators. As outlined in Section \ref{sec:PQ},  there can be two potentially competing contributions to the axion VEV:
\bea
\bar\theta_{\rm PQ}\equiv \frac{\langle a\rangle}{f_a} =\bar\theta_{\rm UV} +\bar\theta_{\rm BSM}, \label{thetapq1}\eea
where $\bar\theta_{\rm UV}$ is the axion VEV induced by
PQ-breaking {\it other than} the QCD anomaly, e.g. the one in Eq. (\ref{thetaUV}) which is induced by
quantum gravity instantons at UV scales, while $\bar\theta_{\rm BSM}$ arises from a combined effect of the PQ-breaking by the QCD anomaly and BSM CP-violation as in Eq. (\ref{theta_bsm}). In our case, $\bar\theta_{\rm UV}$ is essentially a free parameter whose size is characterizing the quality of the PQ symmetry, while $\bar\theta_{\rm BSM}$ from the gluon and quark CEDMs are given by
\dis{
\bar{\theta}_{\rm BSM} = \frac{m_0^2}{2} \sum_{q=u, d, s} \frac{\tilde{d}_q}{m_q} + {\cal O} (4\pi f_\pi^2 w),\label{thetapq}
}
where the piece from the quark CEDMs is calculated with QCD sum rules \cite{Pospelov:2000bw} at the matching scale $\mu=1$ GeV, while the piece  from the gluon CEDM is estimated with NDA at the different matching scale\footnote{For the matching scale of NDA, in this paper
we simply use $\mu_*\simeq 225$ MeV as in  \cite{Weinberg:1989dx}. Note that this is just a matter of choice as there is no systematic way to estimate the uncertainty of NDA. }
$\mu_*\simeq 225$ MeV.
Replacing $\bar{\theta}$ in Eq. (\ref{thepn}) with $\bar{\theta}_{\rm PQ}$, we obtain
\dis{
\Theta_p^{\rm PQ}(\bar{\theta}_{\rm UV}, \tilde{d}_q, d_q) =\,&  \chi m_*  (4 e_u - e_d)  \bar{\theta}_{\rm UV} -  \left(\frac{1}{8}(2\kappa+\xi) + \frac12 \chi m_0^2\right) (4e_u \tilde{d}_u -e_d \tilde{d}_d) + (4d_u-d_d),   \\
\Theta_n^{\rm PQ}(\bar{\theta}_{\rm UV}, \tilde{d}_q, d_q) =\,&  \chi m_*  (4 e_d - e_u)  \bar{\theta}_{\rm UV} + \left(\frac{1}{8}(2\kappa+\xi) + \frac12 \chi m_0^2\right) (4e_d \tilde{d}_d -e_u \tilde{d}_u)  + (4d_d-d_u),
 \label{thepnpq}
}
where notably the strange quark CEDM $\tilde{d}_s$ contribution is cancelled \cite{Pospelov:2000bw}, and the gluon CEDM contribution via $\bar{\theta}_{\rm PQ}$ is ignored, since it is actually negligible compared to the direct contribution in Eq. (\ref{NEDMw}) due to the chiral suppression $(\sim m_*/4\pi f_\pi)$.  
Numerically we then find
\dis{
d_p^{\rm PQ}(\bar{\theta}_{\rm UV}, \tilde{d}_q, d_q, w) =\,& -0.46 \times 10^{-16} \bar{\theta}_{\rm UV} \,e\, \textrm{cm} 
   - e \left( 0.58  \tilde{d}_u +0.073  \tilde{d}_d \right) \\
  & + 0.36  d_u -0.089  d_d -{18} w\, e\,\textrm{MeV}, \\
d_n^{\rm PQ}(\bar{\theta}_{\rm UV}, \tilde{d}_q, d_q, w)  =\,&  0.31 \times 10^{-16} \bar{\theta}_{\rm UV} \,e\, \textrm{cm}  +
  e \left( 0.15  \tilde{d}_u +0.29  \tilde{d}_d \right) \\
 & -0.089  d_u +0.36 d_d + {20} w\,e \,\textrm{MeV}, \label{NEDMpq} 
}
for the nucleon EDMs in the presence of the PQ mechanism.

As mentioned in the previous section, in this work we focus on gauge and/or Higgs-mediated CPV from a new physics sector  which is charged under the QCD gauge group. In this case, typically the quark EDMs $d_q$ are subdominant compared with the quark CEDMs $\tilde{d}_q$. Thus we will neglect the contributions from the quark EDMs in what follows.  In this class of models, moreover, the quark chirality violation in the quark CEDM operators is from the SM Yukawa couplings.
It implies that
\dis{
 \tilde{d}_q(\mu) = m_q K_2 (\mu) \label{GHcond}
} 
with the flavor-independent running coefficient $K_2(\mu)$ as defined in Eq. (\ref{C123}). 
By this relation Eq. (\ref{NEDM}) becomes
\dis{
d_p(\bar{\theta}, K_2, w) &= -0.46 \times 10^{-16} \bar{\theta} \,e\, \textrm{cm}  +  1.1K_2 \,e\, \textrm{MeV} -{18} w\, e\,\textrm{MeV}, \\
d_n(\bar{\theta}, K_2, w) &=  0.31 \times 10^{-16} \bar{\theta} \,e\, \textrm{cm}  -  0.15K_2 \,e\, \textrm{MeV} +{20} w\, e\,\textrm{MeV}
 \label{NEDMGH} 
}
for  $K_2$ and $w$ defined  at $\mu=1$ GeV.
On the other hand, if the PQ mechanism is working for resolving the strong CP problem, Eq. (\ref{NEDMpq}) yields
\dis{
d_p^{\rm PQ}(\bar{\theta}_{\rm UV}, K_2, w) &=  -0.46 \times 10^{-16} \bar{\theta}_{\rm UV} \,e\, \textrm{cm}-1.7 K_2 \,e\, \textrm{MeV} -{18} w\, e\,\textrm{MeV}, \\
d_n^{\rm PQ}(\bar{\theta}_{\rm UV}, K_2, w) &=  0.31 \times 10^{-16} \bar{\theta}_{\rm UV} \,e\, \textrm{cm} +1.7 K_2 \,e\, \textrm{MeV} +{20} w\, e\,\textrm{MeV}
 \label{NEDMGHpq} 
}
again for  $K_2$ and $w$ defined  at $\mu=1$ GeV. Note that here $d_N^{\rm PQ}(K_2)$ includes the contribution from $\bar\theta_{\rm BSM}$ induced by the quark CEDM $\tilde d_q=m_q K_2$.

From the numerical values in Eq. (\ref{NEDMGH}) and Eq. (\ref{NEDMGHpq}), one can observe that
\bea
d_p(\bar{\theta}, w) &\approx& -d_n(\bar{\theta}, w), \label{appr1}\\
d_p^{\rm PQ} (\bar{\theta}_{\rm UV}, K_2, w) &\approx& -d_n^{\rm PQ} (\bar{\theta}_{\rm UV}, K_2, w), \label{appr2}
\eea
unless there is a significant cancellation among the different contributions,
while 
\dis{
d_p(K_2) \approx -7 d_n (K_2). \label{appr3}
}
These approximate relations can be confirmed more precisely by the analytic results Eq. (\ref{thepn}), Eq. (\ref{dN_w}), and Eq. (\ref{thepnpq}) from QCD sum rules. 
Imposing the relation Eq. (\ref{GHcond}) in Eq. (\ref{thepn}) and Eq. (\ref{thepnpq}), we find\footnote{As can be noticed from Eq.(\ref{C2r}),  the magnitude of $d_p(K_2)/d_n(K_2)$  is significantly larger than the unity  partly because of a cancellation among different contributions to $d_n(K_2)$. Note that  $d_N(K_2)$ ($N=p,n$) involve the two isospin-violating parameters $e_u-e_d$ and $\tilde d_u-\tilde d_d$ (see Eq. (\ref{thepn})), therefore there is no reason that $d_p(K_2)/d_n(K_2)$ is close to the unity. }
\bea
\frac{d_p (\bar{\theta})}{d_n(\bar{\theta})}&=& \frac{d_p^{\rm PQ} (\bar{\theta}_{\rm UV})}{d_n^{\rm PQ} (\bar{\theta}_{\rm UV})}  \simeq \frac{4 e_u - e_d}{4 e_d - e_u} = -\frac32, \label{ther}\\
\frac{d_p (K_2)}{d_n(K_2)} &\simeq& - \frac{2\kappa + \xi - 3\chi m_0^2}{2\kappa +\xi - 0.7\chi m_0^2} \simeq -7(1), \label{C2r}\\
\frac{d_p^{\rm PQ} (K_2)}{d_n^{\rm PQ} (K_2)} &\simeq& \frac{4 e_u m_u - e_d m_d}{4 e_d m_d - e_u m_u} \simeq -1. \label{C2pqr}
\eea
where we have used $m_d/m_u \simeq 2$. On the other hand, using Eq. (\ref{dN_w}) and the result of \cite{Yamanaka:2020kjo}, we get
\dis{
\frac{d_p(w)}{d_n(w)} \simeq \frac{d_p^{\rm PQ} (w)}{d_n^{\rm PQ} (w)} \simeq -0.89(2).\label{wr}
}

Given other estimates of the nucleon EDMs based on chiral perturbation theory \cite{Crewther:1979pi,Hockings:2005cn, Mereghetti:2010kp}, lattice-QCD calculations \cite{Dragos:2019oxn, Guo:2015tla, Abramczyk:2017oxr, Alexandrou:2020mds, Bhattacharya:2021lol}, and their discrepancy with the sum rule results, one may trust the above ratios derived from QCD sum rules up to about 50\% uncertainty. 
Then Eqs. (\ref{ther})-(\ref{wr}) confirm the approximate relations in Eqs. (\ref{appr1})-(\ref{appr3}) up to such an uncertainty.

\begin{figure}[h]
    \includegraphics[scale=0.58]{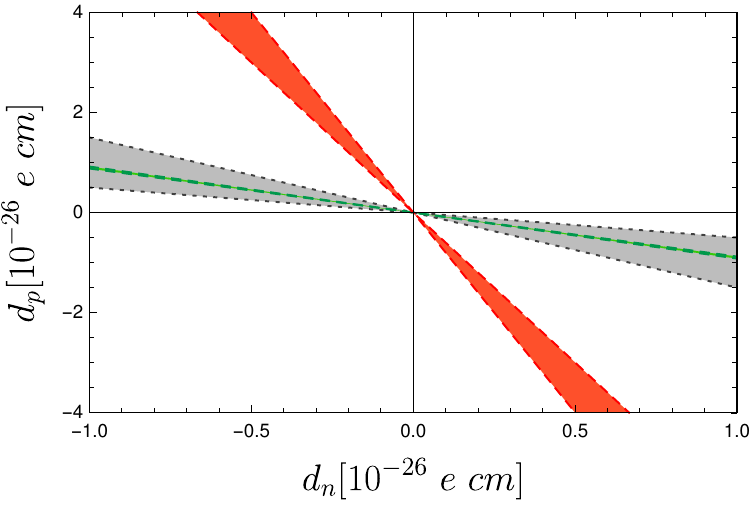}
    \hspace{0.1cm}
    \includegraphics[scale=0.58]{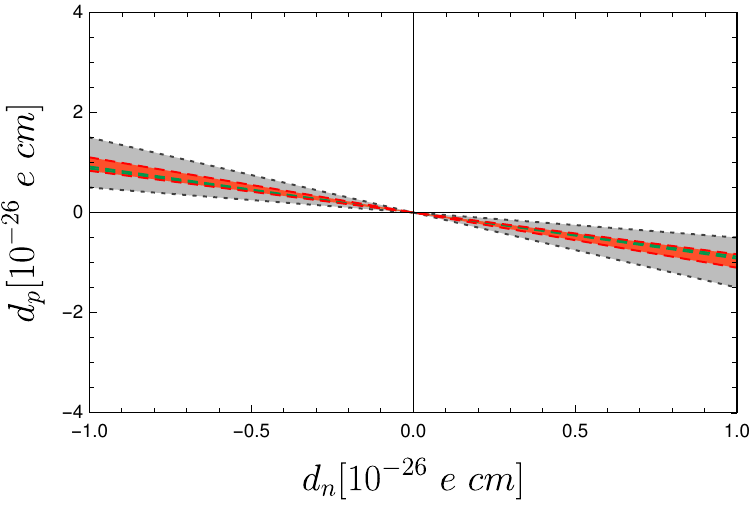}
  \caption{
  The predicted ratios of the proton EDM to the neutron EDM depending on the different origins of CP violation without (left) or with (right) the PQ mechanism. The shaded regions denote the cases where the nucleon EDMs originate dominantly from i) the QCD $\bar{\theta}$-parameter (gray), ii) the quark CEDMs (red), and iii) the gluon CEDM (i.e. the Weinberg operator) (green). 
  Here we assume that the CEDMs are generated at the BSM scale $\Lambda=1$ TeV and subsequently follow the RGE down to the low energy scales. However our results are not sensitive to the value of $\Lambda$.
  }
\label{fig:nEDMs}
\end{figure}

Based on  the above discussion and the RG evolution of CEDMs discussed in the previous section, in Fig. \ref{fig:nEDMs} we depict the nucleon EDM ratio $d_p/d_n$ in three different scenarios that  the dominant source of EDMs  is  i) $\bar\theta$,  or ii) the quark CEDM given by  $\tilde d_q=m_q K_2$ at the BSM scale $\Lambda$ near TeV, or iii) the gluon CEDM at $\Lambda$, either without (left) or with (right) the PQ mechanism. We note that our result reproduces the figures in \cite{deVries:2018mgf, deVries:2021sxz}, while we include the gluon CEDM-dominance as well as the cases without the PQ mechanism beyond the previous works. Our results show that the nucleon EDMs can discern only the scenario ii) without the PQ mechanism, i.e.
the quark CEDM-dominated CP violation without  QCD axion, from the other scenarios.\footnote{The BSM CP violating operators responsible for  the CEDMs can give  quadratically \cite{deVries:2018mgf, deVries:2021sxz} or logarithmically \cite{Morozov:1984goy, Chang:1991hz} divergent radiative correction to
 $\bar{\theta}$, which is likely to provide the dominant source of EDMs \emph{unless} it is cancelled by the PQ mechanism. 
 For this reason, \cite{deVries:2018mgf, deVries:2021sxz} exclude the scenarios of BSM CP violation \emph{without} the PQ mechanism, in which
 $\bar\theta$ is small enough to be subdominant compared to the CEDMs. Here we include these scenarios as an open  possibility and examine their experimental testability by EDM data.  One may simply fine tune $\bar\theta$ to  be small enough even in the presence of BSM CP violation, or there may exist unknown UV mechanism yielding small $\bar\theta$ together with certain form of BSM CP violations.} In the presence of the PQ mechanism, $d_p \approx -d_n$ in all three scenarios i), ii) and iii). Note that EDMs in the scenarios ii) and iii) with the PQ mechanism  include the contribution from the axion vacuum value $\bar\theta_{\rm BSM}$ induced by CEDMs, and  the scenario i) with the PQ mechanism corresponds to the case that the dominant source of EDMs is $\bar\theta_{\rm UV}$ induced by  UV-originated PQ breaking such as quantum gravity effects.
Our results also imply that the nucleon EDMs can not tell us about the origin of non-zero axion vacuum value, for instance they can not discriminate $\bar\theta_{\rm UV}$ from $\bar\theta_{\rm BSM}$.
Thus, we need to look for other CPV observables beyond the nucleon EDMs in order to get information on the quality of the PQ symmetry. In the next subsection, we will discuss the use of some nuclei or atomic EDMs for this purpose.

\subsection{Nuclei or Atomic EDMs}

In the previous subsection, we have noticed that the nucleon EDMs provide only a limited information on the dominant source of CPV and the origin of the axion vacuum value.  In this subsection, we examine whether some nuclei or atomic EDMs are capable of improving the situation.

It has been known that certain nuclei or atomic EDMs are sensitive to CP-odd nuclear forces like the CPV pion-nucleon couplings (see e.g. \cite{1303.2371}). 
Important examples include light nuclei such as $D$ (deuteron) and $^{3} \textrm{He}^{++}$ (helion) whose EDMs might be measured by the storage ring method \cite{Chupp:2017rkp}. The EDMs of diamagnetic heavy atoms like $^{225} \textrm{Ra}$ and $^{129} \textrm{Xe}$ are also such examples \cite{deVries:2021sxz, Osamura:2022rak}. {The EDMs of polar molecules such as ThO, HfF$^+$, and BaF can also be sensitive to the CPV pion-nucleon couplings via the CPV electron-nucleon interaction
$\bar e\gamma_5 e\bar NN$ mediated by the neutral pion \cite{1912.13129, deVries:2021sxz}. Then they may play a role in disentangling the sources of CP violation in the future, although the current sensitivity is not yet competitive to other nuclei or atomic EDMs in terms of hadronic CP violation. Therefore, in this work we will focus on nuclei and diamagnetic atomic EDMs.}

We use the results of \cite{Bsaisou:2014zwa} for the EDMs of  $D$ and $^{3} \textrm{He}^{++}$, yielding 
\bea
\hskip -1cm 
d_D &=& 0.94(1)(d_n+d_p) + \Big[ 0.18(2)\bar{g}_1 {- 0.75(14) \Delta_\pi} \Big] \,e \,\textrm{fm}, \label{dD} \\
\hskip -1cm d_{\rm He} &=& 0.9 d_n-{0.03(1)}d_p \nonumber \\
&&\hskip -0.5cm + \Big[{0.11(1)}\bar{g}_0+ {0.14(2)}\bar{g}_1 {-0.63(15) \Delta_\pi} {-\left(0.04(2) C_1 -0.09(2) C_2 \right) \textrm{fm}^{-3}} \Big] \,e \,\textrm{fm},  \label{dHe} 
\eea
where
 $\bar{g}_{0,1},  \Delta_\pi$, and $C_{1,2}$ are the CPV couplings of pions and nucleons defined as 
\dis{
&
\bar{g}_0 \bar{N} \vec{\sigma} \cdot \vec{\pi} N+ \bar{g}_1 \pi_3 \bar{N} N + m_N \Delta_\pi \pi_3 \vec{\pi}\cdot \vec{\pi}\nonumber  \\
&\hskip -0.5cm +C_1 \bar{N} N D_\mu(N^\dagger S^\mu N) + C_2 \bar{N} \vec{\sigma} N \cdot D_\mu (N^\dagger \vec{\sigma} S^\mu N).
}
Here $\vec{\sigma}=(\sigma_1, \sigma_2, \sigma_3)$ denotes the Pauli matrices for the isospin, $N = (p,n)^T$ is the isospin-doublet nucleon field, and $\vec{\pi}=(\pi_1,\pi_2,\pi_3)$ is the isospin-triplet pion field. We note that the deuteron EDM $d_D$ is not sensitive to $\bar{g}_0, C_1$ and $C_2$ due to the spin and isospin structure of the deuteron \cite{Bsaisou:2014zwa}.

We also consider the atomic EDMs of heavy nuclei $^{225} \textrm{Ra}$ and $^{129} \textrm{Xe}$, using the following results of \cite{deVries:2021sxz, 1807.09581} for $^{225} \textrm{Ra}$ and \cite{Osamura:2022rak,Yanase:2020agg} for $^{129} \textrm{Xe}$:
\bea
d_{\rm Ra} &=& 7.7\times10^{-4}\left[(2.5\pm7.5)\bar{g}_0- (65\pm40)\bar{g}_1 {-(1.1(3.3) C_1-3.2(2.1) C_2) \,\textrm{fm}^{-3} } \right] \,e \,\textrm{fm}, \label{dRa} \nonumber \\ \\
d_{\rm Xe} &=& 1.3\times10^{-5} d_n -10^{-5} \left[1.6\bar{g}_0+ 1.7\bar{g}_1\right] \,e \,\textrm{fm}. \label{dXe}
\eea
The Radium EDM $d_{\rm Ra}$ is sensitive to CPV nuclear forces with relatively weak dependence on the nucleon EDMs due to its octupole deformation \cite{Auerbach:1996zd, Engel:1999np, 1807.09581}. However, its dependence on $\Delta_\pi$ is not currently well known. Also for the Xenon EDM $d_{\rm Xe}$, unfortunately the associated nuclear matrix elements for the contributions from $C_{1,2}$ and  $\Delta_\pi$ are not available at the moment. In what follows, we will discuss 
possible implications of the above results, and also examine
how important the currently unavailable matrix elements are for EDM analysis.

\subsubsection{CPV pion-nucleon couplings 
$\bar{g}_0$ and $\bar{g}_1$
}

The CPV pion-nucleon couplings $\bar{g}_0$ and  $\bar{g}_1$ have been computed by various methods including QCD sum rules and chiral perturbation theory. 
The contributions from the QCD $\bar{\theta}$-parameter were estimated in \cite{deVries:2015una, Bsaisou:2014zwa} using chiral symmetry relation between CPV pion-nucleon couplings 
and quark mass corrections to baryon masses. It results in
\bea
\bar{g}_0 (\bar{\theta})  &=& \frac{\delta m_N}{2f_\pi} \frac{1-\epsilon^2}{2\epsilon} \bar{\theta}  = (15.7\pm1.7)\times10^{-3} \bar{\theta}, \label{gbar0_th}\\
\bar{g}_1 (\bar{\theta}) &=&\left(8c_1 m_N  \frac{\epsilon(1-\epsilon^2)}{16f_\pi m_N} \frac{m_\pi^4}{m_K^2 - m_\pi^2} + {\cal O}\left(\epsilon \frac{m_\pi^4}{m_N^3 f_\pi}\right)\right) \bar{\theta}= -(3.4\pm2.4)\times10^{-3} \bar{\theta}, \nonumber \\  \label{gbar1_th}
\eea
where $\delta m_N = m_n - m_p = 2.49(17)$ MeV, $\epsilon = (m_d-m_u)/2\bar{m} = 0.37(3)$,
$c_1 = 1.0(3) \textrm{GeV}^{-1}$ is related to the nucleon sigma term \cite{Baru:2011bw} as $\sigma_{\pi N} = - 4c_1 m_\pi^2 + {\cal O}(m_\pi^3)$ \cite{Bernard:1996gq}, $f_\pi=92.2$ MeV, $m_K= 495$ MeV, and $m_\pi = 135$ MeV. 
Here $\bar{g}_1$ is subject to large theoretical uncertainty, since the Next-to-Leading Order (NLO) correction is as large as the Leading Order (LO) contribution, and it is uncertain how fast the convergence of the estimation from chiral perturbation theory is. 
In the presence of the PQ mechanism, $\bar\theta$ in Eq. (\ref{gbar0_th}) and Eq. (\ref{gbar1_th}) is replaced by
 $\bar\theta_{\rm PQ}=\langle a\rangle/f_a$  in Eq. (\ref{thetapq1}),  which is determined by the two independent contributions, 
 $\bar\theta_{\rm UV}$  induced by UV-originated PQ breaking such as quantum gravity effects and $\bar\theta_{\rm BSM}$  that originates from the light quark or gluon CEDMs at low energy scales as in Eq. (\ref{thetapq}).

The contributions from the quark CEDMs to the CPV pion-nucleon couplings were estimated with QCD sum rules in \cite{Pospelov:2001ys}, and recently the estimation was improved in \cite{deVries:2021sxz} using chiral symmetry relation and neglecting contribution related to the matrix elements of quark chromomagnetic dipole moments based on the argument of \cite{Seng:2018wwp}. Using the result of \cite{deVries:2021sxz}, we get
\bea
\bar{g}_0 (\tilde{d}_q) &\simeq&  \frac{m_0^2}{8f_\pi}\left[\delta_{g_0} \frac{d \delta m_N}{d \bar{m} \epsilon}(\tilde{d}_u+\tilde{d}_d) -\frac{(1-\epsilon^2)\delta m_N}{\epsilon}\sum_{q=u,d,s}\frac{\tilde d_q}{m_q}\right]
\simeq -0.004(5)  K_2 \, {\rm GeV}^2, \nonumber \\ \label{gbar0_dqt}\\
\bar{g}_1 (\tilde{d}_q) &\simeq & \delta_{g_1} \frac{1}{2f_\pi}(\tilde{d}_u-\tilde{d}_d) \frac{m_0^2}{2} \frac{\sigma_{\pi N}}{\bar{m}} = -0.095(31)\, K_2 \, {\rm GeV}^2 , \label{gbar1_dqt} 
\eea
at the matching scale $\mu= 1$ GeV, where we assumed the relation $\tilde d_q=  m_q K_2$ for the final expression. Here $d \delta m_N/d \bar{m} \epsilon \simeq \delta m_N/\bar{m} \epsilon = 2.49(17)\, \textrm{MeV}/\bar{m} \epsilon$ for
$\bar m = (m_u+m_d)/2 =
3.37(8)$ MeV,  $\sigma_{\pi N} =  59.1(35)$ MeV, and $\delta_{g_0, g_1} = (1.0\pm0.3)$ are introduced to account for theoretical uncertainty.

Finally, the gluon CEDM (Weinberg operator) contribution to $\bar{g}_1$ was computed with QCD sum rules and chiral perturbation theory in \cite{Osamura:2022rak} as
\dis{
\bar{g}_1 (w) \simeq \langle 0| {\cal L}_w | \pi^0 \rangle \left(\frac{\sigma_{\pi N}}{f_\pi^2 m_\pi^2}+ \frac{5 g_A^2 m_\pi}{64\pi f_\pi^4} \right)\simeq \pm (2.6\pm1.5) \times 10^{-3} w\, \textrm{GeV}^2, \label{gbar1_w}
}
at the matching scale $\mu = 1$ GeV, where $ {\cal L}_w = \frac{1}{3} w f^{abc} G_{\alpha}^{a \mu}  G_\mu^{b\delta} \widetilde{G}_\delta^{c\alpha}$ is the Weinberg operator, and $g_A=1.27$. Here the sign ambiguity is from the matrix element of the Weinberg operator estimated by QCD sum rules.
On the other hand, to our knowledge, there has been no dedicated study of $\bar{g}_0$ originating from the gluon CEDM so far. However, the contributions to $\bar{g}_0$ and $\bar{g}_1$ from the gluon CEDM at $\mu = 1$ GeV are expected to be negligible compared to the accompanying  RG-induced quark CEDM contributions.
This can be explicitly seen for $\bar{g}_1(w)$ by applying Eq. (\ref{rgeff}) to Eq. (\ref{gbar1_dqt}) and Eq. (\ref{gbar1_w}), yielding {\bea
\bar{g}_1(\Delta \tilde{d}_q) = {\cal O}(10) \times \bar{g}_1(w), \label{gbar1_delta}\eea
where $\Delta \tilde d_q$ is the RG-induced quark CEDM at $\mu=1$ GeV, which is always accompanying $w$ at 
$\mu=1$ GeV. }
For $\bar{g}_0(w)$, if we use the NDA estimation
\dis{
\bar{g}_0(w) \sim  (m_u+ m_d){\cal O}(4\pi f_\pi w), \label{gbar0_w}
}
it is somewhat bigger than $\bar{g}_0(\Delta \tilde{d}_q)$ obtained from Eq. (\ref{gbar0_dqt}), but still significantly smaller than $\bar{g}_1(\Delta \tilde{d}_q)$.

{We note that the QCD sum rule results for $\bar g_0(\bar\theta)$ and $\bar g_0(\tilde d_q)$ in Eqs. (\ref{gbar0_th}) and (\ref{gbar0_dqt}) are compatible with the NDA estimation implying $\bar g_0(\bar\theta)\sim 4\pi m_* \bar\theta/\Lambda_\chi$ and  $\bar g_0(\tilde d_q)\sim (\tilde d_u+\tilde d_d)\Lambda_\chi$ for the matching scale $\mu_*\simeq 225$ MeV, where $\Lambda_\chi= 4\pi f_\pi$. 
The QCD sum rule result for $\bar g_1(w)$ in Eq. (\ref{gbar1_w}) also is compatible with the NDA estimation implying $\bar g_1(w)\sim (m_u-m_d)\Lambda_\chi w$.
On the other hand, the QCD sum rule results for $\bar g_1(\bar\theta)$ and  $\bar g_1(\tilde d_q)$  in Eqs. (\ref{gbar1_th}) and   (\ref{gbar1_dqt}) are about one order of magnitude bigger than the NDA estimation implying
 $\bar g_1(\bar\theta)\sim 4\pi m_* (m_u-m_d)\bar\theta/\Lambda_\chi^2 $ and  $\bar g_1(\tilde d_q)\sim (\tilde d_u-\tilde d_d)\Lambda_\chi$. This is partly due to that $\sigma_{\pi N}=(m_u+m_d)\langle N|\bar uu +\bar dd|N\rangle/4\simeq 59$ MeV is significantly larger than $m_u+m_d$  at $\mu_*\simeq 225$ MeV. 
 As we will see, such large values of $\bar g_1(\bar\theta)$ and $\bar g_1(\tilde d_q)$ play an important role in our subsequent analysis.

The nuclei and atomic EDMs 
in Eqs. (\ref{dD})-(\ref{dXe}) have better or comparable sensitivity on $\bar{g}_1$ compared to $\bar{g}_0$,
and $\bar{g}_1(\tilde{d}_q)$ is predicted to be an order of magnitude larger than $\bar{g}_0(\tilde{d}_q)$ in Eq. (\ref{gbar0_dqt}) and Eq. (\ref{gbar1_dqt}). 
Eqs. (\ref{gbar0_dqt})-(\ref{gbar0_w}) also imply that $\bar g_1(\Delta \tilde d_q)$ for the RG-induced $\Delta \tilde d_q$ at $\mu=1$ GeV is about an order of magnitude larger than $\bar g_{0,1}(w)$ for $w$ at $\mu=1$ GeV, 
 unless $\bar{g}_0(w)$ is unreasonably bigger than the NDA estimation Eq. (\ref{gbar0_w}).
We can then ignore the contributions from $\bar{g}_{0,1}(w)$ and $\bar{g}_0(\Delta \tilde d_q)$ while focusing only on the contribution
from  $\bar g_1(\Delta d_q)$, 
when we  estimate the EDMs  that originate from the gluon CEDM generated at the BSM scale $\Lambda$.

With the above observations, let us consider 
the ratio $\bar{g}_1/m_n d_n$ which may have certain characteristic values depending on CPV origins.
Assuming the relation $\tilde d_q=m_q K_2$ which is valid for the gauge and Higgs mediated CPV, we find
\bea
\frac{e\bar{g}_1(\bar{\theta})}{m_n d_n(\bar{\theta})} &=& \frac{e\bar{g}_1^{\rm PQ}(\bar{\theta}_{\rm UV})}{m_n d_n^{\rm PQ}(\bar{\theta}_{\rm UV})} \approx -(2.3\pm2.1), \label{r1}\\
\frac{e\bar{g}_1(K_2)}{m_n d_n(K_2)} &\approx& (6.6\pm4.8)\times 10^2, \label{r2}\\
\frac{e\bar{g}_1^{\rm PQ}(K_2)}{m_n d_n^{\rm PQ}(K_2)} &\approx& -(72\pm50),\label{r3} \\
\frac{e\bar{g}_1(\Delta K_2, w)}{m_n d_n(\Delta K_2, w)} &\simeq& \frac{e\bar{g}_1^{\rm PQ}(\Delta K_2, w)}{m_n d_n^{\rm PQ}(\Delta K_2, w)} \approx -(5.0\pm3.5)\, r(\Lambda),\label{r4}
\eea
where $K_2 \equiv (\tilde{d}_q/m_q)_{1 \textrm{ GeV}} $, $\Delta K_2(\Lambda) \equiv (\Delta \tilde{d}_q/m_q)_{1 \textrm{ GeV}} $ for the RG-induced quark CEDM  $\Delta \tilde d_q$ that originates  from the gluon CEDM at  $\Lambda$, and $r(\Lambda) \equiv (\Delta K_2(\Lambda)/w)_{1 \textrm{ GeV}} = 0.41$ (for $\Lambda=1$ TeV), 0.53 (for $\Lambda= 10$ TeV) as given in Eq. (\ref{rgeff}). Note that here $\bar g_1^{\rm PQ}(K_2, w)$ and $d_n^{\rm PQ}(K_2, w)$ include the contributions from the axion vacuum value  $\bar\theta_{\rm BSM}$ induced by the CEDMs, i.e.  
\bea
&&
\bar g_1^{\rm PQ}(K_2, w)= \bar g_1(K_2,w) +\bar g_1(\theta_{\rm BSM}(K_2,w)), \nonumber \\
&& d_n^{\rm PQ}(K_2, w)=d_n(K_2,w)+ d_n(\theta_{\rm BSM}(K_2,w)).\nonumber \eea

From Eqs. (\ref{r1})-(\ref{r4}),
we see that the quark CEDM-dominated CPV scenarios predict clearly different value of $\bar g_1/m_nd_n$ from the $\bar{\theta}$-dominant case regardless of the PQ mechanism. Moreover, the predicted central values are quite different depending on whether there is a QCD axion or not, although they are subject to large uncertainties. On the other hand, Eq. (\ref{r4}) shows that the gluon CEDM-dominated CPV at high scale $\Lambda$ predicts similar value of  $\bar g_1/m_nd_n$  as the $\bar{\theta}$-dominant case, again regardless of the PQ mechanism. Thus, it would be still challenging to discriminate the gluon CEDM-dominant scenario from the $\bar{\theta}$-dominant case even via hadronic CPV observables sensitive to the coupling $\bar{g}_1$.
Yet, if we look at some elements such as $^3$He$^{++}$ (Eq. (\ref{dHe})) and $^{129}$Xe (Eq. (\ref{dXe})), which are equally sensitive to $\bar{g}_0$ as well as $\bar{g}_1$, the $\bar{\theta}$-dominant scenario might be distinguishable from the gluon CEDM-dominant case by the relatively large $\bar{g}_0 (\bar{\theta})$ compared to $\bar{g}_1 (\bar{\theta})$.   
\\

\subsubsection{CPV three-pion coupling $\Delta_\pi$ and four-nucleon couplings $C_1, C_2$}

Currently there is no dedicated study for computation of the CPV three-pion coupling $\Delta_\pi$ and the four-nucleon couplings $C_1, C_2$ from CPV sources such as $\bar{\theta}$ and CEDMs. Thus in this analysis we will use the NDA estimation for them to get an idea of their possible impacts on EDM patterns. 
The NDA rules taking into account the chiral symmetry and the isospin symmetry tell us that 
\bea
\Delta_\pi (\bar{\theta}, \tilde{d}_q, w) &\sim& 4\pi \frac{(m_u-m_d) m_*}{\Lambda_\chi^2} \bar{\theta} + \Lambda_\chi (\tilde{d}_u - \tilde{d}_d) + (m_u-m_d) \Lambda_\chi w,\\
C_1 (\bar{\theta}, \tilde{d}_q, w) &\sim& C_2 (\bar{\theta}, \tilde{d}_q, w) \sim (4\pi)^2 \frac{m_*}{\Lambda_\chi^4} \bar{\theta} +  \frac{4\pi}{\Lambda_\chi^2} (\tilde{d}_u+\tilde d_d) + \frac{4\pi}{\Lambda_\chi} w, \label{C1C2}
\eea
where $\Lambda_\chi = 4\pi f_\pi$, $m_* \simeq m_um_d/(m_u+m_d)$, and the matching scale is $\mu_* \simeq 225$ MeV given by $\alpha_s(\mu_*)/4\pi \simeq 1/6$  \cite{Weinberg:1989dx} for which the one loop QCD beta function is comparable to the two loop QCD beta function. In the following, the above NDA relations will be assumed to hold up to the sign.

If $\Delta_\pi$ obeys the above NDA estimation, and also $\bar{g}_1 (\bar{\theta}, \tilde{d}_q)$ are given by the QCD sum rule results Eqs. (\ref{gbar1_th}) and (\ref{gbar1_dqt}),
the contributions from $\Delta_\pi (\bar{\theta}, \tilde{d}_q)$ to $d_D$ and $d_{\rm He}$ in Eq. (\ref{dD}) and Eq. (\ref{dHe})
are negligible compared to the contributions from $\bar{g}_1 (\bar{\theta}, \tilde{d}_q)$. Considering that the contribution of $\bar g_1(w)$ is overridden by
the contribution from
$\bar{g}_1(\Delta \tilde{d}_q)$, $\Delta_\pi(w)$ can also be ignored for the analysis of $d_D$ and $d_{\rm He}$ under the same assumption.
Although it is not known yet how  $d_{\rm Ra}$ and $d_{\rm Xe}$ depend on $\Delta_\pi$, if the sensitivity  to $\Delta_\pi$ is similar to that of $\bar g_1$, $\Delta_\pi$ 
can be similarly  ignored for the analysis of $d_{\rm Ra}$ and $d_{\rm Xe}$. Therefore in the following analysis for EDM patterns we will ignore $\Delta_\pi (\bar{\theta}, \tilde{d}_q, w)$.

On the other hand, 
concerning the four-nucleon couplings $C_1, C_2$,  one can see from Eq. (\ref{dHe}) and Eq. (\ref{dRa}) that $C_{1, 2} \, \textrm{fm}^{-3} \sim 4\pi f_\pi^3 C_{1, 2}$ should  be compared with $\bar{g}_{0, 1}$ in order to  estimate their relative importance for nuclei or atomic EDMs. Then we find that the contributions from $C_{1, 2} (\bar{\theta}, \tilde{d}_q)$ are at least an order of magnitude smaller than the contribution from $\bar{g}_{0, 1} (\bar{\theta}, \tilde{d}_q)$. In contrast, the contributions from $C_{1,2} (w)$ are larger than the ones from $\bar{g}_{0, 1}(w)$ and $\bar g_0(\Delta\tilde d_q)$, but  comparable to that from  $\bar{g}_{1}(\Delta \tilde{d}_q)$. This is because the four-nucleon couplings $C_{1, 2}$ 
are not suppressed by the light quark masses. Therefore $C_{1, 2}$ can be potentially important if the CP violation is mainly sourced by the gluon CEDM generated at the BSM scale $\Lambda$. In the next subsection, we will examine how $C_{1, 2}(w)$ obeying the NDA estimation can affect EDM patterns depending on their (currently unknown) signs.

\subsubsection{Predicted ratios of nuclei and atomic EDMs to the neutron EDM}

In Fig. \ref{fig:D}, Fig. \ref{fig:He}, and Fig. \ref{fig:RaXe}, we depict the ratios of various nuclei or atomic EDMs to the neutron EDM for the BSM CPV scenarios that we are concerned with.
As anticipated, the quark CEDM-dominant scenario at the BSM scale $\Lambda$ with (blue) or without (red) the PQ
mechanism  shows clearly different pattern from the other scenarios, while the difference between the gluon CEDM-dominance (orange or green) at $\Lambda$ and the $\bar{\theta}$-dominance (gray) is less clear. 
As was noticed in the previous subsections,  the deuteron EDM  is determined mainly by the nucleon EDMs and
$\bar g_1(\bar\theta, \tilde d_q, \Delta \tilde d_q)$, both of which have been evaluated by QCD sum rules.  We then use those sum rule results to obtain $d_D/d_n$ depicted in Fig. \ref{fig:D}.  On the other hand, for the helion EDM
 there can be important contributions from the unknown CP-odd
four-nucleon couplings $C_1(w)$ and $C_2(w)$ induced by the gluon CEDM. Therefore in Fig. \ref{fig:He}
 we plot $d_{\rm He}/d_n$ for four possible sign combinations of $C_1(w)$ and $C_2(w)$ while  assuming that the size of those couplings obeys the NDA as in Eq. (\ref{C1C2}).  Our results indicate that at least half of those possible cases can unambiguously disentangle the gluon CEDM-dominance from the $\bar{\theta}$-dominance.  The Xenon EDM
$d_{\rm Xe}$ in Fig. \ref{fig:RaXe} also might be able to distinguish between the gluon CEDM and the $\bar{\theta}$-parameter via its sensitivity on the coupling $\bar{g}_0$,  if the unknown contribution from 
 $C_1(w)$ and $C_2(w)$ is negligible. We simply assume without justification that it is the case, and plot the resulting $d_{\rm Xe}/d_n$ in Fig. \ref{fig:RaXe}.

In the $\bar\theta$-dominant scenario with the PQ mechanism,
 the axion VEV is by definition induced dominantly by UV-originated PQ-breaking other than the QCD anomaly such as quantum gravity effects,  i.e. $\bar\theta_{\rm PQ}\simeq \bar\theta_{\rm UV}$. On the other hand,  in
the gluon or quark CEDM-dominance scenarios, $\bar\theta_{\rm PQ}\simeq \bar\theta_{\rm BSM}$  again by definition.
As $d_{\rm He}$ and $d_{\rm Xe}$ may discriminate the $\bar\theta$-dominance from the gluon or quark CEDM-dominance, regardless of the presence of the PQ mechanism,  future measurements of those EDMs may provide information not only on BSM CP violation, but also on the origin of the axion VEV, so on the quality of the PQ symmetry.

\begin{figure}[h]
\centering  \includegraphics[scale=0.8]{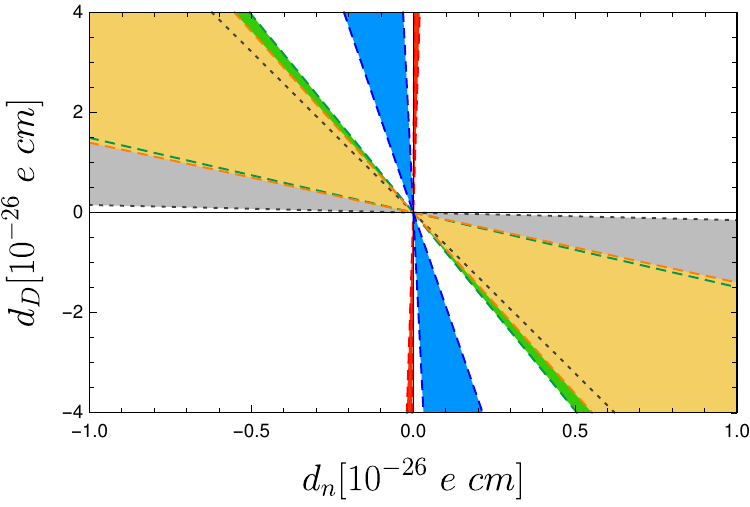} 
  \caption{
  The predicted range of the deuteron EDM compared with the neutron EDM from the CPV sources under consideration.
  The shaded regions denote the cases where the EDMs originate dominantly from i) the QCD $\bar{\theta}$-parameter (gray),  ii) the quark CEDMs without (red) or with (blue) QCD axion, and iii) the gluon CEDM without (green) or with (orange) QCD axion. Here we assume that the CEDMs are generated at $\Lambda=1$ TeV, but the results are not sensitive to the value of $\Lambda$.
  }
\label{fig:D}
\end{figure}

\begin{figure}[h]
    \includegraphics[scale=0.55]{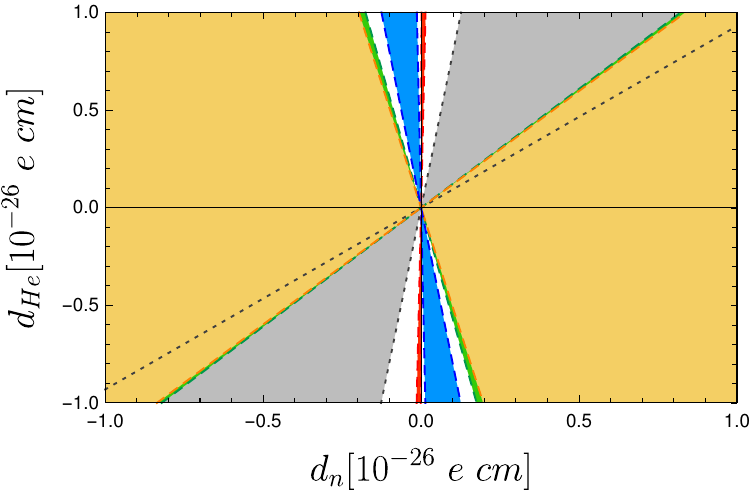}
    \hspace{0.3cm}
    \includegraphics[scale=0.55]{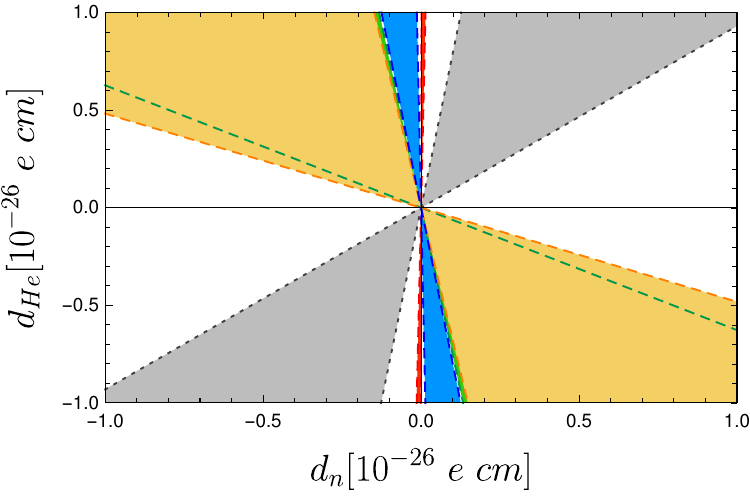} \\
      \includegraphics[scale=0.55]{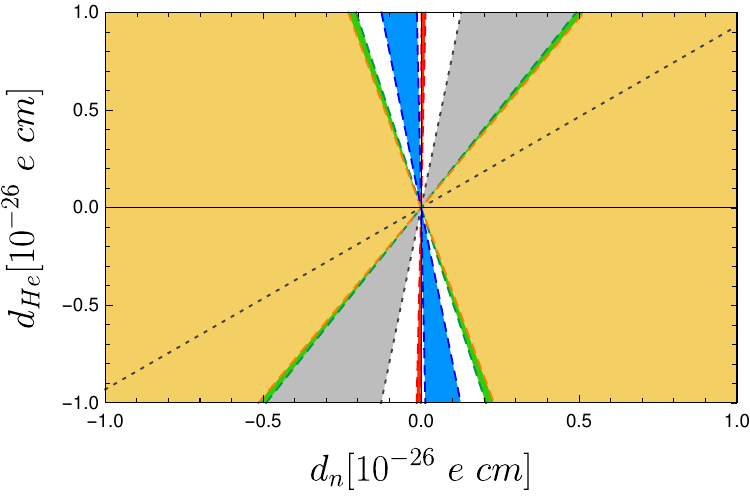}
    \hspace{0.3cm}
    \includegraphics[scale=0.55]{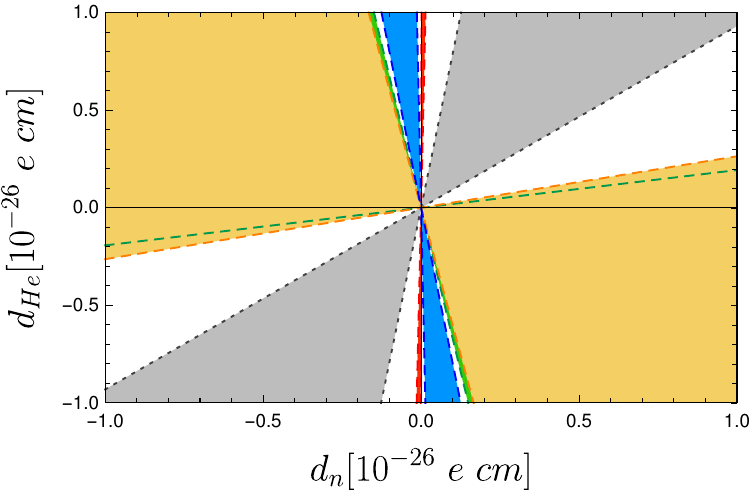}
  \caption{
The predicted range of  $^3$He$^{++}$ EDM compared with the neutron EDM from the CPV sources under consideration. The four plots are obtained by assuming $ s_1 C_1(w)= s_2 C_2(w)= w/f_\pi$ with $(s_1, s_2) = (+1, +1)$ (top-left), $(+1, -1)$ (top-right), $(-1,+1)$ (bottom-left), and $(-1,-1)$ (bottom-right).   
The color code is the same as Fig. \ref{fig:D}: i) the QCD $\bar{\theta}$-parameter (gray),    ii) the quark CEDMs without (red) or with (blue) QCD axion, and iii) the gluon CEDM without (green) or with (orange) QCD axion.  Here we assume that the CEDMs are generated at $\Lambda=1$ TeV, but again the results are not sensitive to the value of $\Lambda$.
  }
\label{fig:He}
\end{figure}

\begin{figure}[h]
  \includegraphics[scale=0.55]{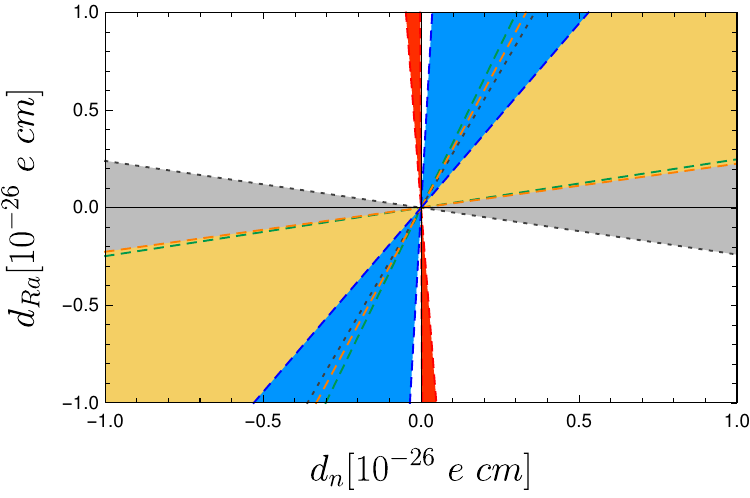} 
   \includegraphics[scale=0.59]{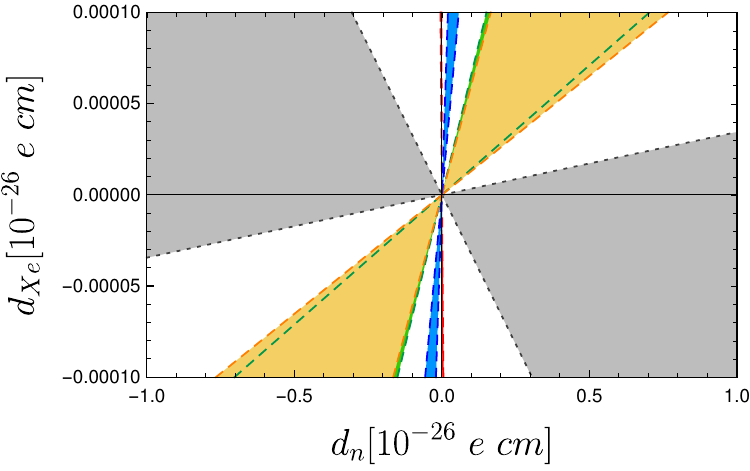} 
  \caption{
 The predicted range of $^{225}$Ra and $^{129} \textrm{Xe}$ EDMs compared with the neutron EDM from the CPV sources under consideration.
The color code is the same as Fig. \ref{fig:D}: i) the QCD $\bar{\theta}$-parameter (gray),    ii) the quark CEDMs without (red) or with (blue) QCD axion, and iii) the gluon CEDM without (green) or with (orange) QCD axion.  The plot for $d_{\rm Xe}$ is obtained by neglecting (potentially important) unknown contributions from CPV four-nucleons contact interactions $C_{1, 2}(w)$, while $d_{\rm Ra}$ turns out to be not so sensitive to them if the NDA estimations for $C_{1, 2}(w)$ are correct. Here we assume that the CEDMs are generated at $\Lambda=1$ TeV, but again the results are not sensitive to the value of $\Lambda$.
  }
\label{fig:RaXe}
\end{figure}

\section{BSM examples} \label{sec:examples}

Here we discuss specific BSM examples which communicate with the SM sector mainly through gauge and Higgs interactions.
As we have discussed in section \ref{sec:GH}, their CP violation will be therefore manifested dominantly via the gluon and quark CEDMs. 

\subsection{Vector-like Quarks}
Vector-Like Quarks (VLQs) may be among the simplest new physics scenarios which transmit CPV to the SM by gauge and Higgs interactions. 
For CP violation, we consider a general renormalizable lagrangian for a VLQ $\psi+\psi^c$ with a real singlet scalar \cite{Choi:2016hro}
\dis{
{\cal L} \supset -\left(m_\psi \psi \psi^c + y_\psi S \psi \psi^c + \textrm{h.c.}\right) -\frac{1}{2} m_S^2 S^2 -A_{SH} S |H|^2 + \cdots ,
}
where the vector-like quark mass $m_\psi$ and the Yukawa coupling $y_\psi$ are complex parameters, and $H$ is the SM doublet Higgs field. 
Here we will discuss this model in some details, because it has not been comprehensively studied before concerning its EDM signatures beyond the scope of \cite{Choi:2016hro}.

 One can remove the phase of the fermion mass by chiral rotation so that a complex CP phase appears in the Yukawa coupling only. Then we may write the lagrangian without loss of generality as
\dis{
{\cal L} \supset -\left(m_\psi \bar{\Psi} \Psi + y_\psi \cos \alpha \,S \bar{\Psi} \Psi + y_\psi \sin \alpha \, S \bar{\Psi} i\gamma_5 \Psi \right)   -\frac{1}{2} m_S^2 S^2 -A_{SH} S |H|^2 + \cdots ,
}
where the parameters $m_\psi$ and $y_\psi$ are now real, and $\alpha$ denotes the CP phase. Here $\Psi \equiv (\psi ~ \psi^{c*})^{T}$ is the Dirac field of the VLQ. If $\cos \alpha \,\sin \alpha \neq 0$ (i.e. $\alpha \neq 0, \pi/2$), CP has to be broken, because $S$ couples to both the CP-even fermion bilinear and the CP-odd 
fermion bilinear.

Assuming the VLQ and the singlet scalar are heavier than the electroweak scale, one can integrate them out. The effective lagrangian below the mass scales of the VLQ and the singlet scalar is then given by some of the operators in Eq. (\ref{eft1}) from the first two diagrams in Fig. \ref{fig:VLQs}.
\dis{
{\cal L}_{\rm CPV} (\mu = \Lambda)=&c_{\widetilde G} f^{abc} G_{\alpha}^{a \mu}  G_\mu^{b\delta} \widetilde{G}_\delta^{c\alpha}+ c_{\widetilde W}  \epsilon^{abc} W_{\alpha}^{a \mu}  W_\mu^{b\delta} \widetilde{W}_\delta^{c\alpha} \\
&+ |H|^2 \left( c_{H\widetilde{G}} G^a_{\mu \nu} \widetilde{G}^{a \mu \nu} + c_{H\widetilde{W}}   W^a_{\mu \nu} \widetilde{W}^{a \mu \nu} +c_{H\widetilde{B}} B_{\mu \nu} \widetilde{B}^{\mu \nu}\right) 
}
with \cite{Weinberg:1989dx, Djouadi:2005gj}
\bea
c_{\widetilde X} &=& -  \frac{1}{12} \frac{g_X^3}{(4\pi)^4} \frac{y_\psi^2}{m_\psi^2} c_\alpha s_\alpha \, 2\textrm{Tr}(T_X^2(\Psi)) 
 \,h(\tau), \\
c_{H \widetilde X} &=& -\frac{g_X^2}{32\pi^2} \frac{y_\psi}{m_\Psi} s_\alpha \frac{A_{SH}}{m_S^2} \,2\textrm{Tr}(T_X^2(\Psi))\,f(\tau),
\eea
where $X= G, W,$ or $B$, $T_X(\Psi)$ is the representation of $\Psi$ in the gauge group associated with the gauge field $X$, $\tau \equiv m_\psi^2/m_S^2$, and the loop functions $h(\tau)$ and $f(\tau)$ are given by
\bea
h(\tau) &=& 4\tau^2 \int^1_0 dx \int^1_0 dy \frac{x^3y^3 (1-x)}{[\tau x(1-xy) + (1-x)(1-y)]^2}, \\
f(\tau)  &=& -2\tau \int_0^1dx \frac{1}{x} \ln[1-x(1-x)/\tau], \nonumber \\
&=& \begin{cases} 
-\tau \left[\ln \left(\frac{1+\sqrt{1-4\tau}}{1-\sqrt{1-4\tau}} \right)-i\pi \right]^2, & \tau < 1/4 \\
4\tau \arcsin^2 (1/2\sqrt{\tau}), & \tau \geq 1/4 
\end{cases}.
\eea
We note that the asymptotic behavior of the loop functions:
\dis{
h (\tau) \simeq \begin{cases} -4\tau \ln \tau & (\tau \ll 1) \\ 1 & (\tau \gg 1) \end{cases}, \quad\quad
f (\tau)  \simeq \begin{cases} \tau^2 & (\tau \ll 1) \\ 1  & (\tau \gg 1) \end{cases}
 }
 
\begin{figure}[t]
\begin{center}
\vspace{-0.3cm}
  \includegraphics[scale=0.45]{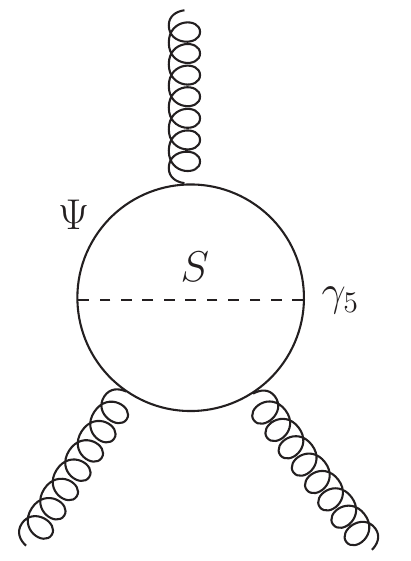}
  \hspace{0.6cm}
    \includegraphics[scale=0.45]{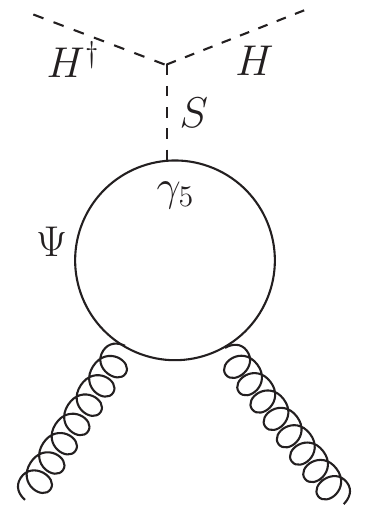}
      \hspace{0.6cm}
    \includegraphics[scale=0.45]{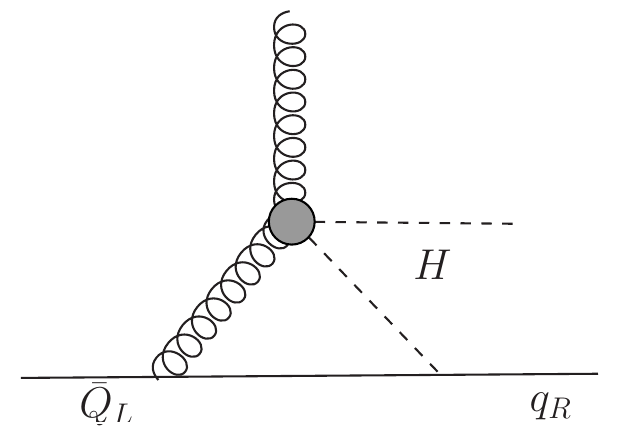}
    \vspace{-0.5cm}
  \end{center}
  \caption{
The diagrams for the dimension-six CPV operators from a VLQ and a singlet scalar. The blob in the third diagram is from the second diagram. If the VLQ is charged under the electroweak gauge groups, the gluons can be replaced by the electroweak gauge bosons.
  }
\label{fig:VLQs}
\end{figure}

The RG equations Eq. (\ref{RGuv}) tells us that the operators in Eq. (\ref{eft1-1}) are also induced at low energies by RG mixing through the third diagram in Fig. \ref{fig:VLQs}, and consequently around the weak scale the following operators in Eq. (\ref{eft2}) are generated\footnote{If the VLQ $\Psi$ is charged under the electromagnetism $U(1)_{\rm em}$, the electron EDM is also generated, which we are not concerned with here.}
\dis{
{\cal L}_{\rm CPV} (\mu = m_W) = \frac{1}{3!} w f^{abc} \epsilon^{\alpha \beta \gamma \delta} G^a_{\mu \alpha} G^b_{\beta \gamma} G^{c\mu}_{\delta}
-\frac{i}{2} \sum_q \left(\tilde{d}_q g_s \bar{q} \sigma^{\mu \nu} G_{\mu \nu} \gamma_5 q+d_q e \bar{q} \sigma^{\mu \nu} F_{\mu \nu} \gamma_5 q  \right).
}
The sizes of the Wilson coefficients are roughly 
\dis{
w \sim  \frac{g_s^3}{(4\pi)^4} \frac{y_\psi^2}{\Lambda^2} s_{2\alpha}, ~ \tilde{d}_q \sim \frac{g_s^2}{(4\pi)^4} \frac{y_\psi}{\Lambda} \frac{m_q}{v} s_\alpha s_\xi, \quad d_q \sim \frac{e^2}{(4\pi)^4} \frac{y_\psi}{\Lambda} \frac{m_q}{v} s_\alpha s_\xi
}
where $\Lambda \sim m_\psi \sim m_S$ and $\xi$ is the Higgs-singlet scalar mixing angle $s_\xi \sim A_SH v/m_S^2$. Therefore, the quark EDMs are relatively small compared with the quark CEDMs by the factor $\alpha/\alpha_s$, and the quark EDMs' contribution to the nuclear and atomic EDMs can be neglected. 

\begin{figure}[h]
\begin{center}
 \begin{tabular}{l}
    \includegraphics[scale=0.45]{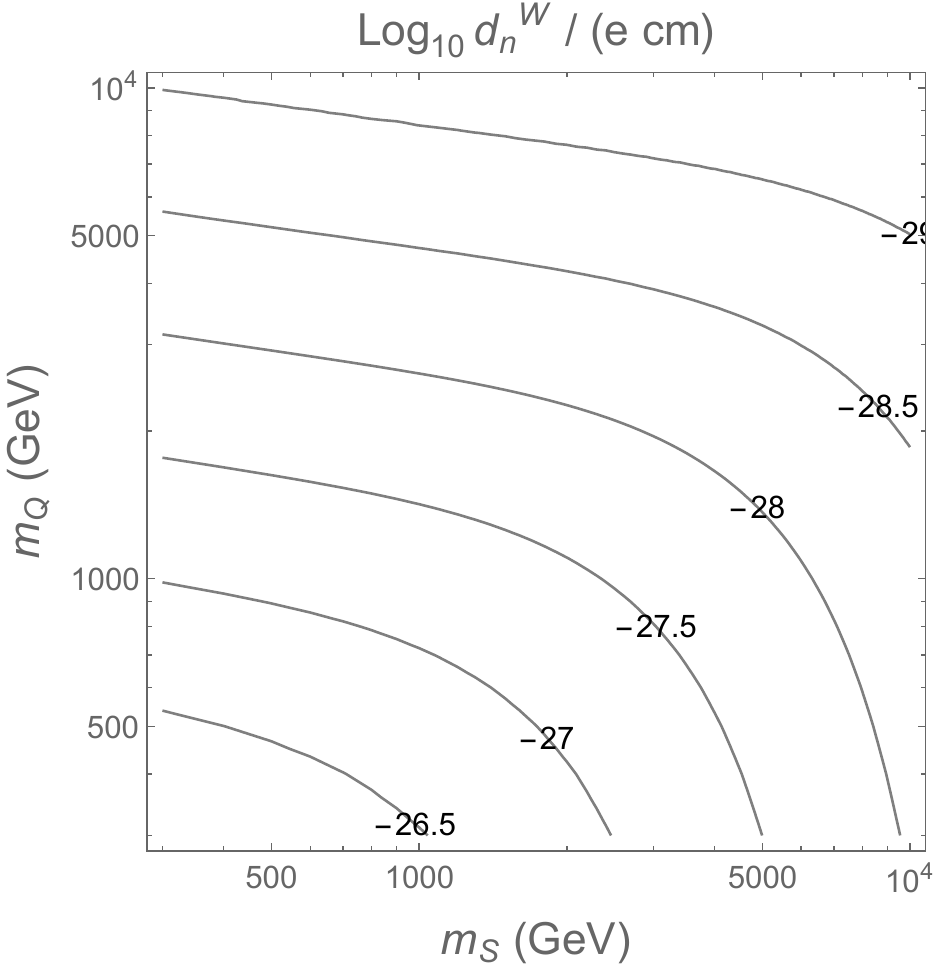}
    \hspace{0.7cm}
 \includegraphics[scale=0.45]{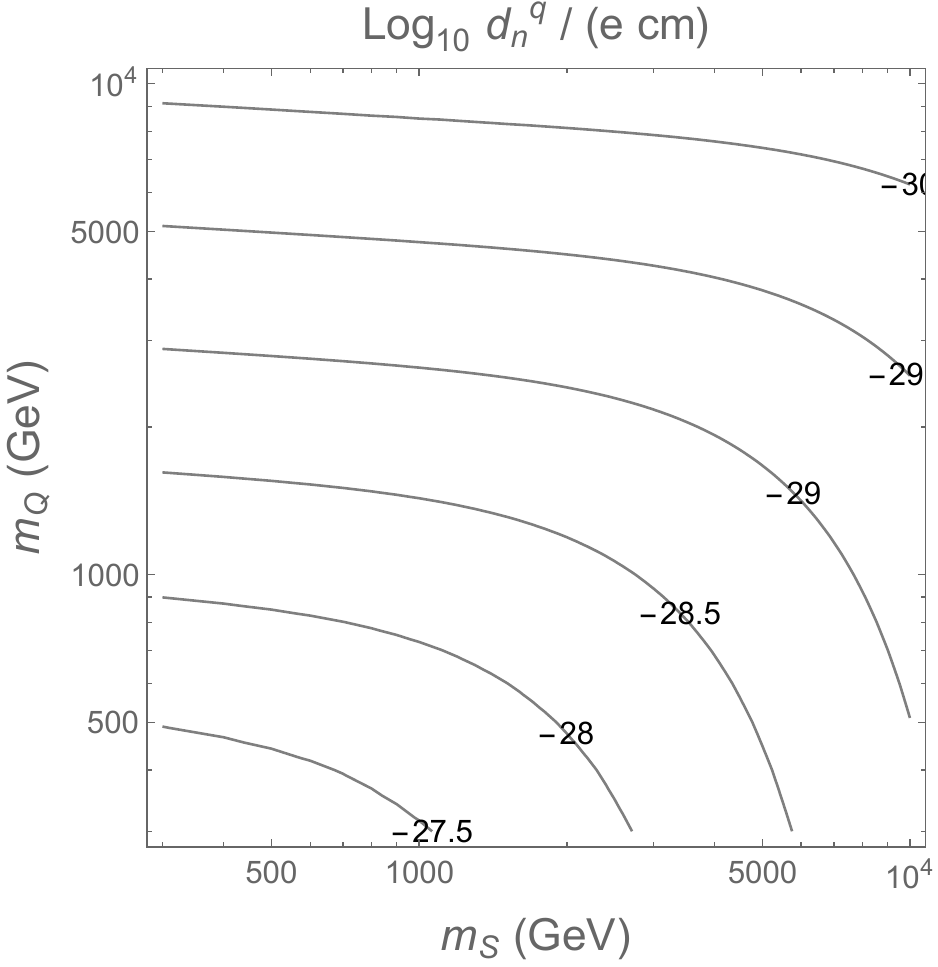}
   \end{tabular}
  \end{center}
  \caption{
The neutron EDM from a CPV VLQ by the Weinberg operator (left) and the RGE induced quark CEDMs (right) with vanishing singlet scalar-Higgs mixing ($\xi=0$). For the plot, we choose
the Yukawa coupling $y_\psi =1$ and the CP angle $\alpha=1$. The Weinberg operator gives a dominant contribution to the neutron EDM for the vanishing mixing angle. 
  }
\label{fig:dn_w}
\end{figure}

In Fig. \ref{fig:dn_w}, we estimate the neutron EDM from the CPV VLQs in terms of VLQ mass $m_Q$ and singlet scalar mass $m_S$ assuming the CP angle $\alpha =1$, no $S$-$H$ mixing ($\xi=0$), and the Yukawa coupling $y_\psi=1$. Even without $S$-$H$ mixing, non-zero quark CEDMs are induced by the RGE from the Weinberg operator as can be seen from Eq. (\ref{analytic}). However, the figure shows that the neutron EDM is dominantly given by the Weinberg operator with about 10$\%$ correction from the RGE-induced quark CEDMs.

In Fig.  \ref{fig:dn_q}, on the other hand, we consider a non-vanishing $S$-$H$ mixing $\sin \xi \simeq v/m_S$ for which sizable quark CEDMs are generated at the UV scale $\Lambda = \min (m_\psi, m_S)$. For this case, the corrections from the RGE are not important for neutron EDM, and the neutron EDM is mostly determined by the quark CEDMs in viable parameter space with $d_n < 10^{-26}\, e\, \textrm{cm}$. The contribution from the Weinberg operator is rather small below 5$\%$.

\begin{figure}[h]
\begin{center}
 \begin{tabular}{l}
    \includegraphics[scale=0.45]{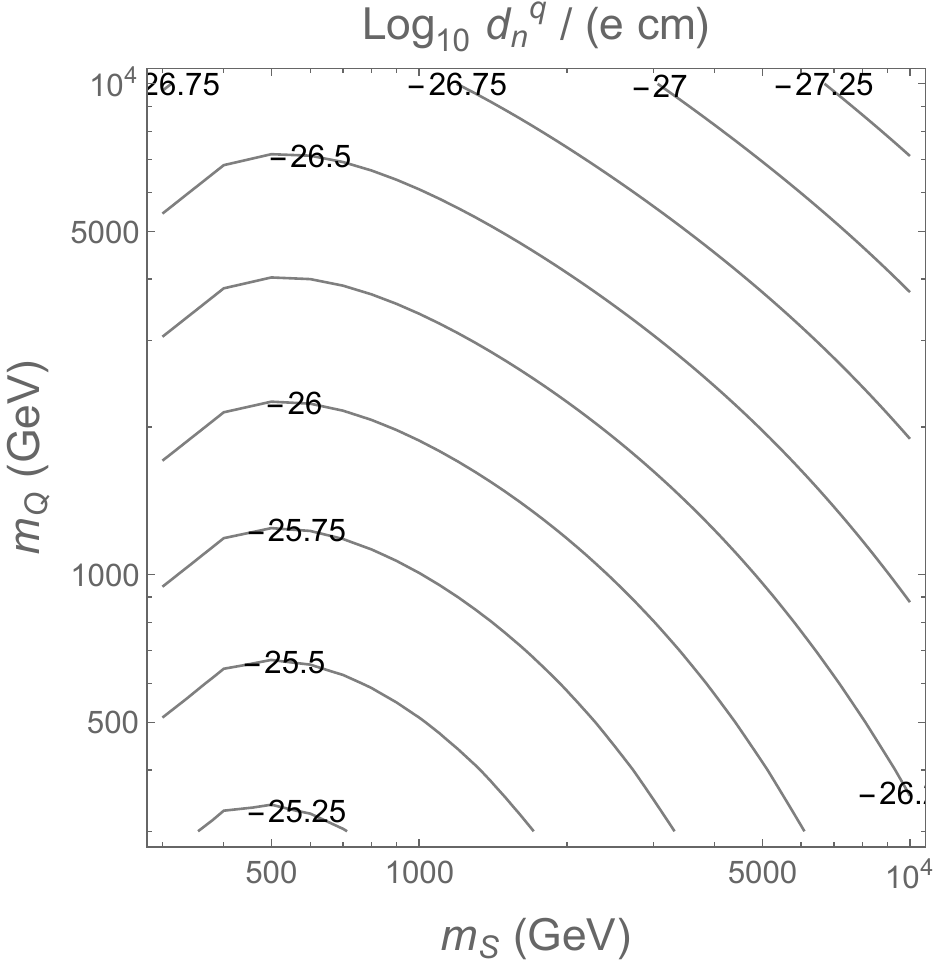}
    \hspace{0.7cm}
 \includegraphics[scale=0.45]{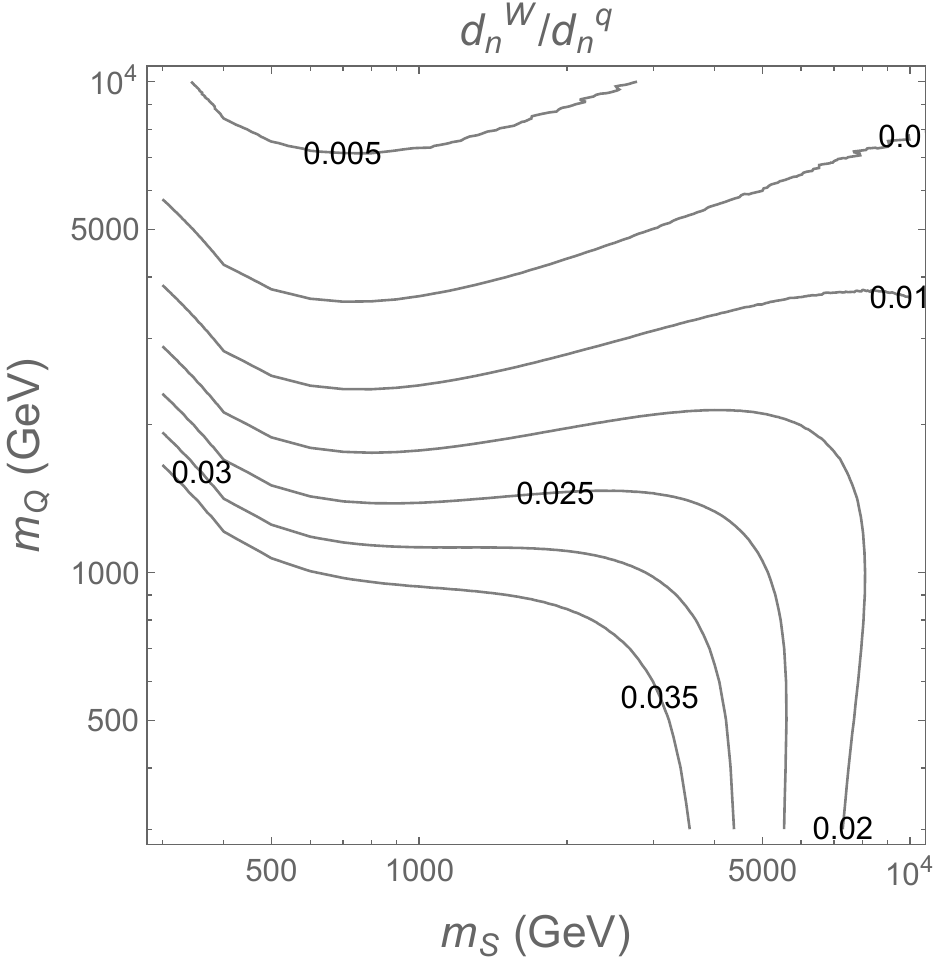}
   \end{tabular}
  \end{center}
  \caption{
The neutron EDM from a CPV VLQ by the quark CEDMs (left) and the ratio of the neutron EDM from the Weinberg operator to the one from the quark CEDMs (right) with non-vanishing singlet scalar-Higgs mixing  $\sin \xi \simeq v/m_S$. For the plot, we choose
the Yukawa coupling $y_\psi =1$ and the CP angle $\alpha=1$. The quark CEDMs give a dominant contribution to the neutron EDM for the non-vanishing mixing angle. 
  }
\label{fig:dn_q}
\end{figure}

\subsection{Supersymmetry}
In supersymmetric (SUSY) extensions of the SM, the dominant CP violating operator is determined by details of the mass spectrum of SUSY particles.

Even in the simplest phenomenologically viable scenarios, such as the MSSM, there are multiple new sources of CPV, which can have a significant impact on the phenomenology of the model.
In the case that sfermions are as light as the gauginos and Higgsinos, the leading CPV operator is typically the quark CEDM \cite{Abel:2001vy} generated by the 1-loop diagram such as the one shown on the right side of Fig. \ref{fig:SUSY}.
CPV is generated by the complex nature of the SUSY breaking parameters, as typically many of them contain a non-zero phase that remains even after performing field redefinitions in gaugino and Higgsino masses.
Other complex parameters of the MSSM include, e.g., squark or slepton mass matrices and bilinear or trilinear couplings;\footnote{We do not discuss them further because they are either subdominant or do not lead to the form of EDM operators we study.} for extensive discussion of these terms, we refer to \cite{Abel:2001vy,Pospelov:2005pr,Nakai:2016atk}.

In fact, these one-loop diagrams involving CPV complex parameters are enhanced by a potentially large $\tan\beta$. 
This can easily lead in a generic SUSY scenario to an electron or neutron EDM that is much larger than experiments allow. The discrepancy between such theoretical expectation and experimental results is called the SUSY CP problem and several explanations for it have been investigated in the literature, an overview of some of them can be found in \cite{Abel:2001vy}.

An apparent solution to evade these constraints is to assume that some SUSY particles are very heavy or that the CPV phases are aligned or canceled by other effects. 
Another, more complete, possibility is to consider specific scenarios of SUSY breaking that achieve this by some well-motivated mechanism, such as split SUSY \cite{Arkani-Hamed:2004ymt,Giudice:2004tc,Arkani-Hamed:2004zhs} or natural SUSY \cite{Dimopoulos:1995mi,Pomarol:1995xc,Cohen:1996vb}.

The former scenario assumes that the scalar superpartners are much heavier than the fermionic ones, such as gauginos and higgsinos. 
This can suppress the EDMs from one-loop diagrams involving scalars, but it also enhances the EDMs from two-loop diagrams involving gauginos.
For example, the gluino can induce a large EDM for the quarks through its interaction with the gluon.
In fact, the split (or high scale) SUSY is an excellent example in which the SUSY CPV is dominantly mediated by gauge and Higgs interactions with the SM sector \cite{Giudice:2005rz,Hisano:2015rna}. 
In particular, the gluon CEDM shown on the left of Fig. \ref{fig:SUSY} can be the dominant CPV operator if the gluino has a mass comparable to that of charginos and neutralinos \cite{Hisano:2015rna}.

On the other hand, the natural SUSY is a scenario where only the superpartners that are relevant for electroweak symmetry breaking, such as stops and higgsinos, are light. 
Such spectrum typically avoids problems associated with fine-tuning, while at the same time it introduces new sources of CPV from the Higgs sector.
For example, a new tree-level interaction between the Higgs and a singlet field (introduced, e.g., to solve the so-called $\mu$ problem) can generate a large EDM for the electron or quarks through two-loop Barr-Zee type diagrams \cite{Nakai:2016atk}.

The extended Higgs sector of the MSSM - which is required to cancel the chiral anomalies - is another source of SUSY contributions to EDMs. 
It consists of two Higgs doublets, which result in five physical Higgs bosons: two CP-even scalars $h$, $H$, one CP-odd pseudoscalar $A$, and two charged scalars $H^\pm$. 
The exchange of these Higgs bosons at one-loop level can induce EDMs for quarks and leptons through their Yukawa couplings and their CKM matrix elements. 
In fact, this type of Higgs sector is a special case of the more general class of models known as type II Two-Higgs-Doublet Models (2HDMs) that predict such extended scalar sector; we discuss them in the next section. 
The EDMs from the extended Higgs sector of the MSSM depend on the masses and couplings of the Higgs bosons, as well as the CPV phase in the Higgs potential. 

Another possibility for SUSY contributions to EDMs is the R-parity violating (RPV) MSSM, which allows for lepton and baryon number violating interactions among the superpartners. 
The RPV MSSM does not introduce new one-loop diagrams contributing to the EDMs \cite{Yamanaka:2012ep}, and the leading contribution takes place at the two-loop level, mainly through the Barr-Zee type diagrams, which involve a loop of charged particles and a loop of neutral particles. 
However, the discussion of the RPV MSSM is beyond the scope of this work, while an extensive discussion can be found in \cite{Yamanaka:2013pfn}.

\begin{figure}[h]
\begin{center}
  \includegraphics[scale=0.42]{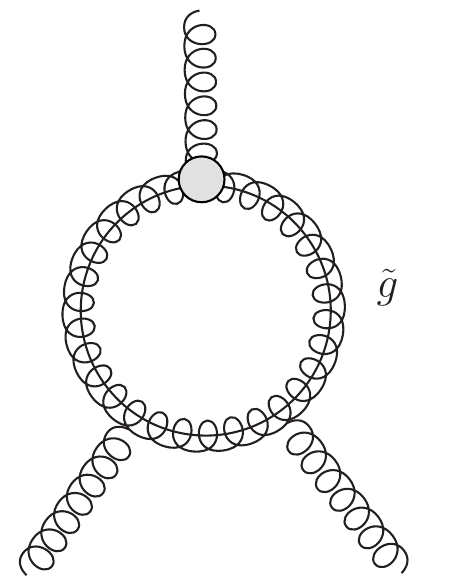}
  \hspace{1cm}
    \includegraphics[scale=0.45]{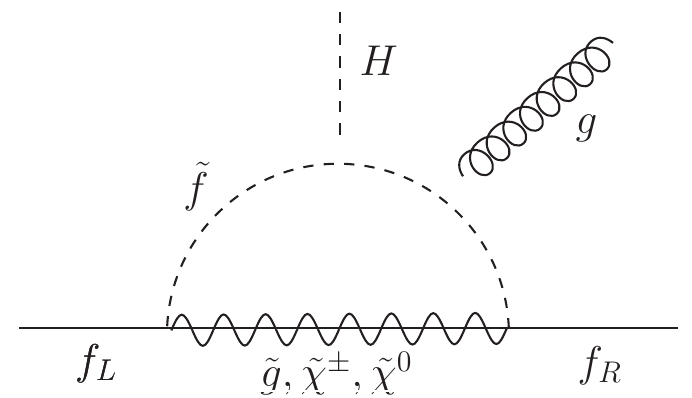}
    \vspace{-0.3cm}
  \end{center}
  \caption{
  The diagrams illustrating the dimension-six CPV operators generated in supersymmetric extensions of the SM. 
  The blob in the first diagram denotes the gluino CEDM originating from the CP phase of gluino mass.
}
\label{fig:SUSY}
\end{figure}

\subsection{2HDMs}
2HDMs are a class of models that can mediate CP violation through heavy beyond the Standard Model (BSM) Higgs bosons, 3 neutral and 2 charged ones, with a $Z_2$ symmetry imposed to suppress the flavor-changing neutral currents, see \cite{Jung:2013hka,Ilisie:2015tra} for an extended discussion of its EDM phenomenology.
CPV phases can enter through both Yukawa interactions, parameterized in general by arbitrary complex matrices,\footnote{The special case of phases described by a scalar matrix corresponds to the so-called Aligned 2HDM.} and by the CPV terms in the potential of neutral scalars.

Compared to the Higgs sector of the MSSM, the 2HDM can potentially exhibit more significant CPV effects, due to the possible presence of physical CP-violating phases in the Higgs sector. 
These CPV phases can exist even if all the input parameters are real and, in contrast to the MSSM, cannot be rotated away by field redefinitions, owing to the absence of R-symmetry. 
Thus, even if the input parameters are chosen to be real, spontaneous symmetry breaking in the 2HDM can give rise to CPV, which does not hold for MSSM at the tree level.
On the other hand, in the MSSM, CP violation can arise from the complex phases of the soft SUSY-breaking parameters or from loop-level effects, as discussed in the previous section, even if the Higgs sector parameters are chosen to be real.

2HDMs are characterized by a rich EDM phenomenology, which depends largely on how the Higgs doublets couple to the SM fermions, and therefore fall into several types - see, e.g., \cite{Branco:2011iw} for an overview.
In these models, the quark CEDMs are the dominant CPV operators, and they can be generated by the top quark loops, as illustrated in Fig. \ref{fig:2HDMs}, which also involves the exchange of neutral and charged Higgs bosons. 
Another significant source of CPV emerges from the CEDM of the gluon \cite{Jung:2013hka}.
In contrast, the CPV four-fermion operators, which arise from the exchange of two heavy Higgs bosons, are typically negligible.
This takes place because they are suppressed by the product of two small Yukawa couplings and the absence of the potentially large factor $\tan^3\beta$; the parameter $\tan\beta$ is the ratio of the vacuum expectation values of the two Higgs doublets, which determines the strength of the Yukawa couplings.
Therefore, the EDMs in 2HDMs with a $Z_2$ symmetry are mainly sensitive to the quark and gluon CEDMs.

\begin{figure}[h]
\begin{center}
  \includegraphics[scale=0.45]{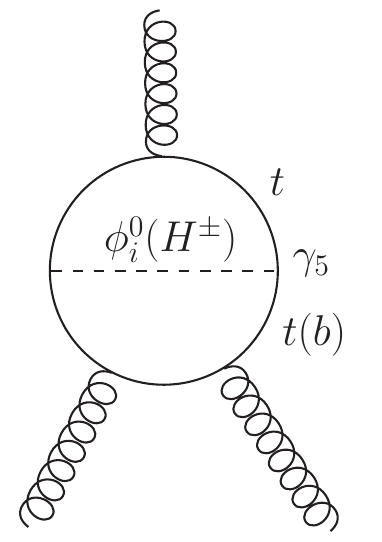}
  \hspace{1cm}
    \includegraphics[scale=0.42]{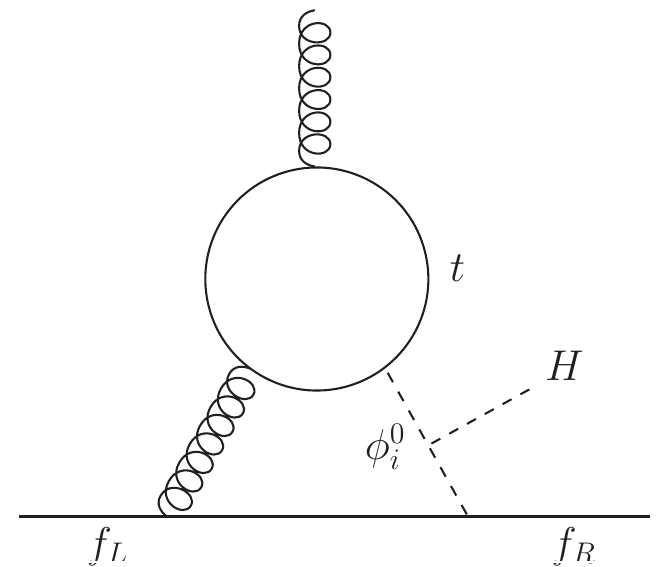}
    \vspace{-0.3cm}
  \end{center}
  \caption{
  The diagrams depict the dimension-six CPV operators originating from 2HDMs. Here, $\phi_i^0 = h, H^0, A^0$ denotes the neutral Higgs bosons. 
  The left panel illustrates the generation of the Weinberg operator, while the right one presents the generation of the quark CEDM.
  }
\label{fig:2HDMs}
\end{figure}

\section{Conclusions} \label{sec:conc}

Since the SM predictions of the  nuclear and atomic EDMs from the Kobayashi-Maskawa phase are well below the current and near-future experimental bounds, the observation of non-vanishing EDM in near future indicates that the underlying CP violation is due to the QCD $\theta$-parameter or a BSM source. In this work, we have examined whether future EDM measurements of nucleons and some nuclei/atoms can give us information not only on the origin of CP violation, but also on the PQ mechanism for the dynamical relaxation of the QCD $\theta$-parameter. 

In the presence of the PQ mechanism, BSM CP violation affects EDMs both directly and by shifting  the axion vacuum value when combined with the PQ breaking by the QCD anomaly. On the other hand, PQ breaking other than the QCD anomaly, 
e.g. quantum gravity effects,
which typically takes place  at UV scales and characterizes the quality of the PQ symmetry, affects the EDMs {\it mostly} by shifting the axion vacuum value.
By this reason, the pattern of  EDMs of different elements can be sensitive to the existence of the QCD axion and the quality of the associated PQ symmetry, in addition to providing information on the effective operators describing  BSM CP violation at low energy scales.  

To be concrete and for simplicity, we focus on a class of BSM scenarios where BSM CP violation is dominantly mediated to the SM sector by the SM gauge and Higgs interactions. In this class of BSM scenarios,  flavor-conserving CP violation around the QCD scale may appear dominantly in the form of the gluon and light quark CEDMs  and/or the QCD $\theta$-term. Motivated examples include vector-like quarks and certain parameter spaces of the MSSM and the 2 Higgs-doublet models.

We find that the nucleon EDMs show a distinctive pattern when the EDMs are dominantly induced by the light quark CEDMs without the PQ mechanism. 
In the presence of the PQ mechanism,  the axion vacuum value which determines 
the QCD $\theta$-parameter  might be induced either by  CEDMs or by 
UV-originated PQ breaking other than the QCD anomaly, for instance the PQ breaking by quantum gravity effects. We find that in case with the PQ mechanism the nucleon EDMs  have a similar pattern regardless of what is the dominant source of EDMs among the CEDMs and $\theta$-term, unless there is a significant cancellation between the contributions from different sources.
In contrast, some nuclei or atomic EDMs can have characteristic patterns significantly depending on the dominant source  of EDMs, which may allow identifying the dominant source  of CP violation among the CEDMs and $\theta$-term.
Yet, discriminating the gluon CEDM from the QCD $\theta$-parameter necessitates additional knowledge of low energy parameters induced by the gluon CEDM, which is not available at the moment.
Our results imply that   EDMs can reveal unambiguous sign of BSM CPV while identifying the origin of the axion vacuum value, however it requires further knowledge of low energy parameters  associated with  BSM CPV.
More extensive studies on this matter with additional  BSM CPV sources and hadronic/leptonic CPV observables are subject to future work \cite{preparation}.

\section*{Acknowledgments}
This work was supported by IBS under the project code, IBS-R018-D1. We thank Nodoka Yamanaka for helpful discussions.

\appendix
\section{RGE of the CPV dimension-six operators} \label{app:RGE}

In the gauge and Higgs mediated CP violation, the CPV effect above the electroweak scale appears through the following dimension-six operators of the SM gauge fields and the Higgs field, as given in Eq. (\ref{eft1}) and (\ref{eft1-1}):
\dis{
&{\cal L}_{\rm CPV} = c_{\widetilde G} f^{abc} G_{\alpha}^{a \mu}  G_\mu^{b\delta} \widetilde{G}_\delta^{c\alpha}+ c_{\widetilde W}  \epsilon^{abc} W_{\alpha}^{a \mu}  W_\mu^{b\delta} \widetilde{W}_\delta^{c\alpha} \\
&+ |H|^2 \left( c_{H\widetilde{G}} G^a_{\mu \nu} \widetilde{G}^{a \mu \nu} + c_{H\widetilde{W}}   W^a_{\mu \nu} \widetilde{W}^{a \mu \nu} +c_{H\widetilde{B}} B_{\mu \nu} \widetilde{B}^{\mu \nu}\right) + c_{H\widetilde{W} B} H^\dagger \tau^a H \widetilde{W}^a_{\mu \nu} B^{\mu \nu} \\
&+\left(\sum_{q=u, d}  \sum_{X=G, W, B} i(c_{q X})_{ij} \bar{Q}_{Li} \sigma^{\mu \nu} X_{\mu \nu} q_{Rj} H^{(*)} + \sum_{X=W, B} i(c_{e X})_{ij} \bar{L}_{i} \sigma^{\mu \nu} X_{\mu \nu} e_{Rj} H^{(*)} + \textrm{h.c.} \right) \label{CPVeft}
}
where the Wilson coefficients $c_\alpha \, (\alpha=\widetilde{G}, \widetilde{W}, ...)$ are all real-valued, $i, j$ denotes flavor indices, and $H^{(*)}\equiv H $ or $H^*$ in order to make the operators invariant under the SM gauge groups.

The RG equations of the above dimension-six operators at one-loop are given in \cite{1303.3156, Jenkins:2013zja, Jenkins:2013wua, Alonso:2013hga}. Here we use the Yukawa couplings defined as 
\dis{
{\cal L}_{\rm Yukawa} = -\left[(Y_u)_{ij} \bar{u}_{R i} Q_{L j} H + (Y_d)_{ij} \bar{d}_{R i} Q_{L j} H^* + (Y_e)_{ij} \bar{e}_{R i} L_j H^* + \textrm{h.c.}\right]
}   
with the flavor indices $i, j$. 
The other parameters appearing in the following RG equations are
$c_{A, 3}= N_c$, $c_{A, 2}=2$,  $c_{F,3}=(N_c^2-1)/2N_c$, $c_{F,2} = 3/4$, 
$b_{0, 3} = 11N_c/3-2 n_F/3$, $b_{0,2} = 22/3-1/6-(N_c+1)$, $b_{0,1}= -1/6-(11N_c/9+3)$ with $N_c=3$, and $q_\psi$ denotes the $U(1)_Y$ hypercharge of the field $\psi$.
The RG equations for the operators in Eq. (\ref{CPVeft}) at one-loop are then given by
\bea
16\pi^2\frac{d c_{\widetilde G}}{d \ln \mu} &=&  (12 c_{A, 3} - 3 b_{0, 3}) g_3^2 c_{\widetilde G}\,, \\
16\pi^2 \frac{d c_{\widetilde W}}{d \ln \mu} &=&  (12 c_{A, 2} - 3 b_{0, 2}) g_2^2 c_{\widetilde W}\,,
\eea
%%%
\bea
16\pi^2 \frac{d c_{H\widetilde  G}}{d\ln\mu} &=&\left( -6 q_H^2 g_1^2-\frac92 g_2^2-2b_{0,3}g_3^2 \right) c_{H\widetilde G}+\left( 2 i g_3  \textrm{Tr}[Y_u c_{uG} + Y_d c_{dG}]  + \textrm{h.c.} \right), \nonumber \\ \\
16\pi^2 \frac{d c_{H\widetilde  W}}{d\ln\mu}&=& -15 g_2^3c_{\widetilde W} + \left(-6 q_H^2 g_1^2-\frac52 g_2^2-2b_{0,2}g_2^2 \right) c_{H\widetilde W}  + 2 g_1 g_2 q_H c_{H\widetilde WB}\,, \\
16\pi^2 \frac{d c_{H\widetilde  B}}{d\ln\mu} &=& \left(2 q_H^2  g_1^2-\frac92 g_2^2 -2b_{0,1}g_1^2\right) c_{H\widetilde B} +6 g_1 g_2 q_H c_{H\widetilde WB}\,, \\
16\pi^2 \frac{d c_{H\widetilde  WB}}{d\ln\mu}&=& 6 g_1 g_2^2 q_H c_{\widetilde W} + \left( -2 q_H^2 g_1^2+\frac92 g_2^2-b_{0,1} g_1^2-b_{0,2} g_2^2\right) c_{H\widetilde WB} \nonumber \\
&& + 4 g_1 g_2 q_H c_{H\widetilde B}  + 4 g_1 g_2 q_H c_{H\widetilde W}\,,
\eea
%%%
\dis{
16\pi^2  \frac{d (c_{uG})_{ij}}{d \ln \mu} 
&=  \left[ \left(10 c_{F,3} - 4c_{A,3} -b_{0,3}\right) g_3^2 -3 c_{F,2} g_2^2 + \left( -3 q_u^2+8q_u q_Q -3 q_Q^2 \right) g_1^2\right] (c_{uG})_{ij} \\
&+ 8 c_{F,2} g_2 g_3 (c_{uW})_{ij} +  4 g_1 g_3 (q_u+q_Q) (c_{uB})_{ij} \\
&+\textrm{Im}\left[-4(Y_u^\dagger)_{ij} g_3 (c_{HG} + i c_{H \widetilde G}) +3 g_3^2 c_{A,3} (Y_u^\dagger)_{ij} \left(c_{G} + i c_{\widetilde G} \right)\right], \label{cuG}
}
%%%
\dis{
16\pi^2  \frac{d (c_{uW})_{ij}}{d \ln \mu} 
&=\left[  2 c_{F,3}  g_3^2 + \left( 3 c_{F,2}-b_{0,2} \right) g_2^2 + \left( -3 q_u^2+8q_u q_Q -3 q_Q^2 \right) g_1^2\right] (c_{uW})_{ij} \\
&+ 2 c_{F,3} g_2 g_3 (c_{uG})_{ij} + g_1 g_2 (3 q_Q-q_u) (c_{uB})_{ij} \\
& -\textrm{Im}\left( (Y_u^\dagger)_{ij} \left[ g_2 (c_{HW} + i c_{H \widetilde W})- g_1 (q_Q+q_u)  (c_{HWB} + i c_{H \widetilde WB}) \right]\right),
}
%%%
\dis{
16\pi^2  \frac{d (c_{uB})_{ij}}{d \ln \mu} 
&= \left[ 2 c_{F,3}  g_3^2  -3 c_{F,2} g_2^2 + \left(3 q_u^2+4q_u q_Q +3 q_Q^2 -b_{0,1}\right) g_1^2\right] (c_{uB})_{ij} \\
&+ 4 c_{F,3} g_1 g_3 \left(q_u+q_Q\right) (c_{uG})_{ij} + 4 c_{F,2} g_1 g_2 (3q_Q-q_u) (c_{uW})_{ij} \\
& -\textrm{Im}\left( (Y_u^\dagger)_{ij}\left[2 g_1 (q_Q+q_u)   (c_{HB} + i c_{H \widetilde B}) - \frac32 g_2  (c_{HWB} + i c_{H \widetilde WB}) \right]\right),
}
%%%
%%%
\dis{
16\pi^2  \frac{d (c_{dG})_{ij}}{d \ln \mu} 
&=  \left[ \left(10 c_{F,3} - 4c_{A,3} -b_{0,3}\right) g_3^2 -3 c_{F,2} g_2^2 + \left( -3 q_d^2+8q_d q_Q -3 q_Q^2 \right) g_1^2\right] (c_{dG})_{ij} \\
&+ 8 c_{F,2} g_2 g_3 (c_{dW})_{ij}  +  4 g_1 g_3 (q_d+q_Q) (c_{dB})_{ij} \\ 
& +\textrm{Im}\left[-4(Y_d^\dagger)_{ij} g_3(c_{HG} + i c_{H \widetilde G}) +3 g_3^2 c_{A,3} (Y_d^\dagger)_{ij} \left(c_{G} + i c_{\widetilde G} \right)\right],
}
%%%
\dis{
16\pi^2  \frac{d (c_{dW})_{ij}}{d \ln \mu} 
&=\left[ 2 c_{F,3}  g_3^2 + \left( 3 c_{F,2}-b_{0,2} \right) g_2^2 + \left( -3 q_d^2+8q_d q_Q -3 q_Q^2 \right) g_1^2\right] (c_{dW})_{ij} \\
&+ 2 c_{F,3} g_2 g_3 (c_{dG})_{ij} + g_1 g_2 (3 q_Q-q_d) (c_{dB})_{ij} \\
& -\textrm{Im}\left((Y_d^\dagger)_{ij} \left[ g_2(c_{HW} + i c_{H \widetilde W}) +g_1 (q_Q+q_d)  (c_{HWB} + i c_{H \widetilde WB}) \right]\right),
}
%%%
\dis{
16\pi^2  \frac{d (c_{dB})_{ij}}{d \ln \mu} 
&= \left[ 2 c_{F,3}  g_3^2  -3 c_{F,2} g_2^2 + \left(3 q_d^2+4q_d q_Q +3 q_Q^2 -b_{0,1}\right) g_1^2\right] (c_{dB})_{ij} \\
&+ 4 c_{F,3} g_1 g_3 \left(q_d+q_Q\right) (c_{dG})_{ij} + 4 c_{F,2} g_1 g_2 (3q_Q-q_d) (c_{dW})_{ij} \\
& -\textrm{Im}\left( (Y_d^\dagger)_{ij}\left[2 g_1 (q_Q+q_d)   (c_{HB} + i c_{H \widetilde B}) + \frac32 g_2  (c_{HWB} + i c_{H \widetilde WB}) \right]\right),
}
%%%
%%%
\dis{
16\pi^2  \frac{d (c_{eW})_{ij}}{d \ln \mu} 
&=\left[  \left( 3 c_{F,2}-b_{0,2} \right) g_2^2 + \left( -3 q_{e}^{2}+8q_{e} q_{L} -3 q_L^2 \right) g_1^2\right] (c_{eW})_{ij}  + g_1 g_2 (3 q_L-q_e) (c_{eB})_{ij} \\
& -\textrm{Im}\left((Y_e^\dagger)_{ij} \left[ g_2 (c_{HW} + i c_{H \widetilde W}) +g_1 (q_L+q_e) (c_{HWB} + i c_{H \widetilde WB}) \right]\right),
}
%%%
\dis{
16\pi^2  \frac{d (c_{eB})_{ij}}{d \ln \mu} 
&= \left[   -3 c_{F,2} g_2^2 + \left(3 q_e^2+4q_e q_L +3 q_L^2 -b_{0,1}\right) g_1^2\right] (c_{eB})_{ij}+ 4 c_{F,2} g_1 g_2 (3q_L-q_e) (c_{eW})_{ij} \\
& -\textrm{Im}\left(  (Y_e^\dagger)_{ij} \left[2 g_1 (q_L+q_e)  (c_{HB} + i c_{H \widetilde B}) + \frac32 g_2 (c_{HWB} + i c_{H \widetilde WB}) \right]\right). \label{ceB}
}
%%%
The RG equations for $c_{q X}$ and $c_{eX}$ in Eqs. (\ref{cuG})-(\ref{ceB}) involve the Wilson coefficients of the following CP-even operators through the complex phase of the Yukawa couplings:
\dis{
{\cal L}_{\textrm{CP-even}} &= c_{G} f^{abc} G_{\alpha}^{a \mu}  G_\mu^{b\delta} G_\delta^{c\alpha}+ |H|^2 \left( c_{HG} G^a_{\mu \nu} G^{a \mu \nu} + c_{HW}   W^a_{\mu \nu} W^{a \mu \nu} +c_{HB} B_{\mu \nu} B^{\mu \nu}\right) \\
&+ c_{HW B} H^\dagger \tau^a H W^a_{\mu \nu} B^{\mu \nu}\,. 
}

\bibliography{edm_ctpu}
\bibliographystyle{utphys}

\end{document}